\documentclass[%
 aps,
 prl,
amsmath,amssymb,
reprint,
longbibliography
]{revtex4-1}

\usepackage{graphicx}
\usepackage{dcolumn}
\usepackage{bm}

\usepackage[utf8]{inputenc}
\usepackage{mathptmx}
\usepackage{multirow}
\usepackage{braket}

\usepackage[english]{babel}
\usepackage{xcolor}
\colorlet{RED}{red}
\colorlet{BLUE}{blue}
\usepackage{dcolumn}
\usepackage{bm}
\usepackage[version=4]{mhchem} 
\usepackage{acronym}
\usepackage{adjustbox}
\usepackage{float}
\usepackage{tikz}
\usepackage{microtype} 
\usepackage{algorithm}
\usepackage{xcolor}
\usepackage{algpseudocode}

\usetikzlibrary{calc,shapes.geometric,decorations.pathmorphing,patterns}

\definecolor{background-color}{gray}{0.98}
\usepackage[margin=2.3cm,bmargin=1cm,footnotesep=1cm]{geometry}

\usepackage[utf8]{inputenc}
\usepackage{amsmath}
\usepackage{amssymb}
\usepackage{amsthm}

\begin{document}


\title{
Single-reference coupled-cluster theory based on the multi-purpose cluster operator
}

\author{Karol Kowalski}
\email{karol.kowalski@pnnl.gov}
\affiliation{%
  Physical Sciences Division, 
  Pacific Northwest National Laboratory, Richland, Washington, 99354, USA
}
\affiliation{%
Department of Physics, University of Washington, Seattle, Washington 98195, USA
}

\author{Nicholas P. Bauman}
\affiliation{%
  Physical Sciences Division, 
  Pacific Northwest National Laboratory, Richland, Washington, 99354, USA
}



\date{July 2025}

\begin{abstract} 
In this paper, we develop a theoretical framework that extends single-reference (SR) coupled-cluster (CC) theory beyond its conventional role of describing a single electronic state—typically the lowest-energy state within the symmetry sector defined by the reference determinant. Rather than viewing the SR-CC cluster operator solely as a device for reproducing one target state, we consider more general constructions in which different components of the cluster operator play distinct roles, ranging from encoding states of different symmetry than the reference to enabling SR-CC Ansatz to describe multiple states simultaneously. These developments lead to a new class of SR-CC downfolding formalisms in which the resulting active-space effective Hamiltonians are capable of concurrently representing multiple correlated states  nonorthogonal to the reference function. We establish three theorems that formalize this extension and demonstrate that standard CC downfolding emerges as a special case of the proposed framework. Finally, we introduce a Hermitian variant based on a unitary CC representation, which enables realistic simulations of ground and excited states while reducing the quantum resources required.
\end{abstract}

\maketitle

\section{Introduction}
Coupled-cluster (CC) theory has evolved into a leading quantum many-body framework for describing correlation effects across a wide range of energy scales, with applications spanning nuclear physics, quantum chemistry, and extended condensed-phase systems.
\cite{coester58_421,coester60_477,cizek66_4256,purvis82_1910,paldus72_50,bishop1987coupled,paldus07,Bartlett2007}
Alongside the extensive development and application of standard single-reference (SR) CC methods, a broad class of multi-reference formulations has been introduced to address situations that elude a single-determinant description. Such cases typically involve quasi-degenerate manifolds or multiple low-lying states of the many-body Hamiltonian.  

Over the past several decades, substantial progress has been made in the development of genuine multi-reference coupled-cluster (MR-CC) theories \cite{lyakh2012multireference,sinha2012development,datta2012multireference,musial2019intermediate,aoto2016internally,feldmann2024renormalized,adam2025multireference} and closely related multi-reference many-body perturbation theory (MR-MBPT) approaches.\cite{li2015multireference, sokolov2016time, lechner2021perturbative, li2023intruder, hayashi2024quasi, abraham2024novel,verma2025multireference,king2025bridging} In particular, significant effort has been devoted to constructing intruder-state-free formulations capable of treating systems characterized by diverse and complex patterns of entanglement. Despite these advances, MR-CC and MR-MBPT methods share a common limitation: their algebraic complexity and the associated high computational cost. These challenges typically stem from the need to treat increasingly complex excitation manifolds, which in certain regimes can also give rise to serious numerical instabilities and degraded accuracy. 

It is therefore highly desirable to develop SR-CC-based formulations that can, at least partially, assume the role traditionally played by MR-CC and MR-MBPT methods - namely, the simultaneous description of multiple electronic states—while retaining the relative simplicity and favorable scaling of single-reference approaches. Ideally, such formulations would also provide a systematic and compact route to constructing effective Hamiltonians, analogous to those employed in multi-reference theories.
One area that would particularly benefit from readily accessible effective Hamiltonians is quantum computing. In this context, limited quantum resources can be leveraged to diagonalize effective (or downfolded) Hamiltonians defined in reduced-dimensionality spaces, enabling simulations that would otherwise be infeasible on near-term hardware. Several approaches, such as multi-reference driven similarity group approach,\cite{huang2023leveraging} unitary transformation of Hamiltonian based on exact quartic truncation scheme, \cite{lang2020unitary} dynamical self-energy mapping,\cite{dhawan2021dynamical} transcorrelated Hamiltonians,\cite{kumar2022quantum} and CC downfolding  \cite{kowalski2018properties,bauman2019downfolding,kowalski2021dimensionality} have recently been proposed and explored for this purpose. The importance of these formulations lies in the ability to tune the dimensionality of the relevant (active) spaces to match the capabilities of available quantum hardware. 

To address this challenge, we depart from standard coupled-cluster (CC) formulations, in which the sole role of the cluster operator $T$ is to reproduce an exact or approximate form of the full configuration interaction (FCI) expansion generated by the operator $C_{\rm FCI}$ in the intermediate normalization, i.e.,
\begin{equation}
e^{T} |\Phi\rangle =  C_{\rm FCI} |\Phi\rangle \;,
\label{etfci}
\end{equation}
where $|\Phi\rangle$ denotes the reference function.
In this work, we introduce the concept of a \emph{multi-purpose cluster operator}, whose individual components are designed to serve distinct physical and computational objectives.
We illustrate this concept through examples ranging from symmetry-breaking mechanisms in CC theory to the construction of state-universal effective Hamiltonians derived within a simple single-reference (SR) framework.
These effective Hamiltonians can be used to simultaneously describe ground and excited states that have non-zero overlap with the reference function.
This capability is particularly appealing for the treatment of quasi-degenerate ground and excited states, including excited states characterized by complex correlation effects involving high-rank excitations. In this paper, we analyze both non-Hermitian and Hermitian formulations of state-universal coupled-cluster downfolded Hamiltonians.

\section{Single-reference coupled-cluster theory} 
In the single-reference CC (SR-CC) formulation, which is typically identified with ground-state applications, the correlated wave function $|\Psi\rangle$ is represented through the exponential Ansatz defined by the cluster operator $T$,
\begin{equation}
|\Psi\rangle = e^T |\Phi\rangle \;,
\label{eq1}
\end{equation}
where $|\Phi\rangle$ stands for the reference function represented by a single Slater determinant (usually, chosen as a Hartree--Fock (HF)  Slater determinant). 

The CC energy and cluster amplitudes for a system defined by Hamiltonians $H$ can be evaluated by solving energy-independent CC equations 
\begin{eqnarray}
   Q(e^{-T}He^{T})|\Phi\rangle &=& 0 \;, \label{eq2} \\
   \langle\Phi|e^{-T}He^{T}|\Phi\rangle &=& E \;, \label{eq3}
\end{eqnarray}
where $Q$ stands for the projection operator onto a subspace spanned by excited configurations by acting with $T$ onto $|\Phi\rangle$. For the exact case ($T$ operators include all possible excitations) and for the standard approximations,  equations (\ref{eq2})-(\ref{eq3}) are equivalent at the solution to the energy-dependent form of CC equations:
\begin{equation}
    (P+Q) He^T |\Phi\rangle = E (P+Q)e^T |\Phi\rangle \;,
    \label{eqe}
\end{equation}
where $P=|\Phi\rangle\langle\Phi|$ is the projection operator onto the reference function.
Both representations will be employed in the following analysis.

\section{coupled-cluster downfolding}
The CC formalism plays an important role in the design of various embedding and downfolding approaches.
(see Refs.~\cite{evangelista2014driven,welborn2016bootstrap,huang2023leveraging,shee2024static,feldmann2024complete,weisburn2025multiscale}).
Recently, it was demonstrated that 
the  CC energy, alternatively to the textbook formula (\ref{eq3}), can be obtained as an eigenvalue of the active-space effective Hamiltonian $H^{\rm eff}$
\cite{kowalski2018properties},
\begin{equation}
    H^{\rm eff} = (P+Q_{\rm int}) e^{-T_{\rm ext}} H e^{T_{\rm ext}} (P+Q_{\rm int}) \;,
    \label{eq4}
\end{equation}
where $Q_{\rm int}$ is a projection operator onto all excited Slater determinants in the active space w.r.t. $|\Phi\rangle$. In deriving this result (known as Sub-systems Embedding Sub-algebras (SES) Theorem \cite{kowalski2018properties}) we use the partitioning of the cluster operator $T$ into its component acting in the active-space ($T_{\rm int}$) and component that correlates the active space with entire remaining Hilbert space ($T_{\rm ext}$). The partitioning 
\begin{equation}
    e^T|\Phi\rangle = e^{T_{\rm ext}} e^{T_{\rm int}} |\Phi\rangle \;,
    \label{eq5}
\end{equation} 
was introduced by Adamowicz and Piecuch in the active space CC formulations \cite{pnl93,piecuch1994state}. If $T_{\rm int}$ generates all excited Slater determinants in the active space when acting onto $|\Phi\rangle$ then 
according to the SES Theorem \cite{kowalski2018properties}  we have 
\begin{equation}
H^{\rm eff} e^{T_{\rm int}} |\Phi\rangle = E
e^{T_{\rm int}} |\Phi\rangle \;.
\label{eq6}
\end{equation}
we will further refer to the process of forming an effective Hamiltonian that compresses all correlation effects outside of the active space as CC downfolding.\\
%

\section{coupled-cluster Downfolding based on the broken-symmetry solutions}

The first application of the multi-purpose cluster operator formalism addresses the symmetry-breaking problem in coupled-cluster theory, that is, the ability to encode information about states whose symmetry differs from that of the reference function \( |\Phi\rangle \) in the exponential Ansatz. Although the symmetry-breaking mechanism has previously been examined in the context of homotopy-based analysis of multiple CC solutions for the H$_4$ benchmark system,\cite{jankowski1999physical} the present work extends this mechanism to a general case (for more recent analysis of multiple solutions of CC equations and their practical applications see Refs.~\cite{mayhall2010multiple,lee2019excited,faulstich2023homotopy,faulstich2024coupled,sverrisdottir2024exploring}).

\paragraph{Theorem 1 (General mechanism for symmetry breaking in the single-reference CC theory).---}  In the exact limit, where the cluster operator $T$   
includes all possible excitations with respect to the reference function $|\Phi\rangle$, there exists a symmetry-breaking mechanism that enables single-reference CC theory to describe states whose symmetries differ from that of the reference function.
\paragraph{Proof.---} 
%
%
%
For simplicity, let us consider a situation where the system possesses two types of symmetries, 
${\cal S}_1$ (the symmetry of the reference function) and ${\cal S}_2$, 
defined by the projection operators 
$P + Q_{{\cal S}_1}$ and $Q_{{\cal S}_2}$, respectively. 
Here, $Q_{{\cal S}_1}$ denotes the projection operator onto configurations of ${\cal S}_1$ symmetry 
that are orthogonal to $|\Phi\rangle$. 
We assume that the cluster operator $T$ can be decomposed into components 
$T_{{\cal S}_1}$ and $T_{{\cal S}_2}$ ($T = T_{{\cal S}_1} + T_{{\cal S}_2}$), 
which, when acting on $|\Phi\rangle$, generate all possible excited configurations 
of ${\cal S}_1$ and ${\cal S}_2$ symmetry, respectively.
%
Projecting Eq.~(\ref{eqe}) onto $P+Q_{{\cal S}_1}$ and $Q_{{\cal S}_2}$ and keeping in mind that  $[H,P+Q_{{\cal S}_1}]=[H,Q_{{\cal S}_2}]=0$ we get the following equations:
\begin{widetext}
\begin{eqnarray}
 (P+Q_{{\cal S}_1})H (P+Q_{{\cal S}_1}) e^{T_{{\cal S}_1}+T_{{\cal S}_2}}|\Phi\rangle &=& E
    (P+Q_{{\cal S}_1}) e^{T_{{\cal S}_1}+T_{{\cal S}_2}}|\Phi\rangle \;, \label{eq7} \\
 Q_{{\cal S}_2} H Q_{{\cal S}_2}  e^{T_{{\cal S}_1}+T_{{\cal S}_2}}|\Phi\rangle &=& EQ_{{\cal S}_2}
    e^{T_{{\cal S}_1}+T_{{\cal S}_2}}|\Phi\rangle \;. \label{eq8}
\end{eqnarray}
\end{widetext}
The above equations consist of two eigenvalue problems - the first corresponding to the ${\cal S}_1$ symmetry, while the second corresponds to the ${\cal S}_2$ symmetry. Let us assume that one is interested in energy $E_{{\cal S}_2}$ of the ${\cal S}_2$ state. Then Eq.~(\ref{eq8}) for $T_{{\cal S}_2}$ amplitudes says that 
$Q_{{\cal S}_2}  e^{T_{{\cal S}_1}+T_{{\cal S}_2}}|\Phi\rangle$ is the corresponding  eigenvector, which in the  full configuration interaction (FCI) takes the form $C_{{\cal S}_2}|\Phi\rangle$, i.e., 
\begin{equation}
 Q_{{\cal S}_2}  e^{T_{{\cal S}_1}+T_{{\cal S}_2}}|\Phi\rangle = C_{{\cal S}_2}|\Phi\rangle   \;.  \label{eq10}
\end{equation}
In the following, we will denote the $i$-th many-body components of $T_{{\cal S}_1}$, $T_{{\cal S}_2}$, and 
$C_{{\cal S}_2}$ by $T_{{\cal S}_1,i}$, $T_{{\cal S}_2,i}$, and $C_{{\cal S}_2,i}$ $i=1,\ldots,N_e$ ($N_e$ stands for a total number of correlated electrons). Since $|\Phi\rangle$ is of ${\cal S}_1$ symmetry, then the scalar $C_{{\cal S}_2,0}=0$.
Comparing the excitation ranks of the left- and right-hand sides of  Eq.~(\ref{eq10}) indicates that there is a recursive one-to-one mapping that can be established between $T_{{\cal S}_2}$ and $C_{{\cal S}_2}$
for any form of the $T_{{\cal S}_1}$ operator, i.e., 
\begin{eqnarray}
 T_{{\cal S}_2,1} &=&  C_{{\cal S}_2,1} \;,\label{cr1} \\
 T_{{\cal S}_2,2} &=& C_{{\cal S}_2,2} \nonumber \\
 &&- Q_{{\cal S}_2}(\frac{1}{2}C_{{\cal S}_2,1}^2+C_{{\cal S}_2,1}T_{{\cal S}_1,1}+\frac{1}{2}T_{{\cal S}_1,1}^2 ) \;, \label{cr2}\\
 \ldots &  &  \nonumber
\end{eqnarray} 
If there are no degeneracies between $E_{{\cal S}_2}$ and energies corresponding to the ${\cal S}_1$ symmetry, then the Eq.~(\ref{eq7}) for the $T_{{\cal S}_1}$ operator is no longer eigenvalue problem but rather represents polynomial equations for $T_{{\cal S}_1}$ amplitudes parametrically depending on the $T_{{\cal S}_2}$ ones and energy $E_{{\cal S}_2}$. If the solution  $T_{{\cal S}_1}$ of  Eq.~(\ref{eq7}) exists then the CC energy calculated from the energy expression (\ref{eq3}) equals $E_{{\cal S}_2}$.
\qedsymbol{}

In the above proof, the two components \( T_{{\cal S}_1} \) and \( T_{{\cal S}_2} \) play complementary roles. The operator \( T_{{\cal S}_2} \) is primarily responsible for recovering the eigenvector \( C_{{\cal S}_2}|\Phi\rangle \) [Eq.~(\ref{eq10})], whereas \( T_{{\cal S}_1} \) acts as a ``messenger'' that mediates the coupling between the \( {\cal S}_1 \) and \( {\cal S}_2 \) symmetry sectors.
The SES Theorem for CC formulations based on the ${\cal S}_1$ symmetry reference function assures that for a broken symmetry solutions, disussed in Theorem 1, it is possible to construct effective Hamiltonians in active spaces to reproduce the $E_{{\cal S}_2}$ energy  even in the case when active space contains no ${\cal S}_2$ configurations.

\section{State-universal-type coupled-cluster downfolding: non-Hermitian variant}
Recently, the SES theorem (\ref{eq6}) motivated the development of effective Hamiltonians for quantum computing and machine learning applications targeting the ground-state applications \cite{bauman2019downfolding,liang2024effective,bauman2025coupled}. Although time-dependent extensions of the downfolding formalism \cite{downfolding2020t} may pave the way for excited states extensions, the more elegant approach would be a multi-state, or state-universal, variant of the SR-CC SES theorem. To this end, we will prove the theorem:

\paragraph{Theorem 2 (State-universal CC downfolding  based on a single-reference CC formalism).---} There exists a single-reference type external cluster operator $\Sigma_{\rm ext}$ that defines state-universal effective Hamiltonian, $H^{\rm eff}$, in arbitrary complete active space (CAS),
$H^{\rm eff} = (P+Q_{\rm int}) e^{-\Sigma_{\rm ext}}He^{\Sigma_{\rm ext}}(P+Q_{\rm int})$, that furnishes energies of arbitrary $K$ states having non-zero overlap with the reference function $|\Phi\rangle$, $1\leq K \leq M$, where $M$ is the dimensionality of the CAS. 
\paragraph{Proof.---} 
In the proof of Theorem 2 we utilize the properties of the CASCC parametrization of the exact wave functions, introduced by Adamowicz and collaborators \cite{ivanov2000casccd,ivanov2000new,adamowicz2000new,ivanov2009multireference}, i.e.,
\begin{equation}
    |\Psi\rangle = e^{T_{\rm ext}}C_{\rm int}|\Phi\rangle \;,
    \label{adam1}
\end{equation}
where $T_{\rm ext}$ is the external-type cluster operator (as in Eq.~(\ref{eq5})) and $C_{\rm int}$ is a configuration interaction (CI)  type  operator acting in CAS of interest. We will first prove the validity of Theorem 2 for $K=2$ case. Let us assume that we use two exact CASCC parametrizations:
\begin{eqnarray}
    |\Psi_{\rm A}\rangle &=& e^{T^{\rm (A)}_{\rm ext}} A_{\rm int}|\Phi\rangle \;, \label{tma} \\
    |\Psi_{\rm B}\rangle &=& e^{T^{\rm (B)}_{\rm ext}} B_{\rm int}|\Phi\rangle \;,\label{tmb} 
\end{eqnarray}
where $A_{\rm int}$ and $B_{\rm int}$ are CAS CI operators associated with various electronic states having non-zero overlap with $|\Phi\rangle$.
Instead of using these parametrizations for diagonalizing $H$ directly as in the CASCC formalism, we will utilize the fact that the similarity transformed Hamiltonian $\bar{H}$, $\bar{H} = S^{-1} H S$ has the same spectrum as Hamiltonian $H$, assuming that $S$ is a non-singular operator. In particular, we will use
Ansatz (\ref{tma}) to diagonalize similarity transformed Hamiltonian $\bar{H}_B=e^{-T^{\rm (B)}_{\rm ext}}He^{T^{\rm (B)}_{\rm ext}}$ and Ansatz (\ref{tmb}) to diagonalize similarity transformed Hamiltonian $\bar{H}_A=e^{-T^{\rm (A)}_{\rm ext}}He^{T^{\rm (A)}_{\rm ext}}$, i.e., 
\begin{eqnarray}
e^{-T^{\rm (B)}_{\rm ext}}He^{T^{\rm (B)}_{\rm ext}} e^{T^{\rm (A)}_{\rm ext}} A_{\rm int}|\Phi\rangle &=& E_{\rm A}
e^{T^{\rm (A)}_{\rm ext}} A_{\rm int}|\Phi\rangle \;,\label{sca} \\
e^{-T^{\rm (A)}_{\rm ext}}He^{T^{\rm (A)}_{\rm ext}} e^{T^{\rm (B)}_{\rm ext}} B_{\rm int}|\Phi\rangle &=& E_{\rm B}
e^{T^{\rm (B)}_{\rm ext}} B_{\rm int}|\Phi\rangle \;,\label{scb}
\end{eqnarray}    
where $E_{\rm A}$ and $E_{\rm B}$ are the energies of states $|\Psi_{\rm A}\rangle$ and $|\Psi_{\rm B}\rangle$, respectively.
Premuliplying Eqs. (\ref{sca}) and (\ref{scb}) from the left by 
$e^{-T^{\rm (A)}_{\rm ext}}$ and $e^{-T^{\rm (B)}_{\rm ext}}$, 
introducing cumulative external cluster operator   $\Sigma_{\rm ext}(T^{\rm (A)}_{\rm ext},T^{\rm (B)}_{\rm ext})=T^{\rm (A)}_{\rm ext}+T^{\rm (B)}_{\rm ext}$ 
($[T^{\rm (A)}_{\rm ext},T^{\rm (B)}_{\rm ext}]=0$), and projecting onto CAS defined by $P+Q_{\rm int}$ projection operator, we obtain 
\begin{eqnarray}
    H^{\rm eff} A_{\rm int}|\Phi\rangle &=& E_{\rm A} A_{\rm int}|\Phi\rangle  \;, \label{heffa1} \\
    H^{\rm eff} B_{\rm int}|\Phi\rangle &=& E_{\rm B} B_{\rm int}|\Phi\rangle  \;, \label{heffa1}
\end{eqnarray}
where 
\begin{widetext}
\begin{equation}
H^{\rm eff} = (P+Q_{\rm int}) e^{-\Sigma_{\rm ext}(T^{(\rm A)}_{\rm ext},T^{(\rm B)}_{\rm ext})} H e^{\Sigma_{\rm ext}(T^{(\rm A)}_{\rm ext},T^{(\rm B)}_{\rm ext})} (P+Q_{\rm int}) \;, \label{keq2}
\end{equation}
\end{widetext}
which concludes the proof for $K=2$. The extension of this 
%
%
%
%
proof to arbitrary $K$ follows the same steps as for $K=2$. Let us assume that we have $K$ expansions
\begin{equation}
|\Psi(i)\rangle = e^{T^{(i)}_{\rm ext}} C_{\rm int}^{(i)}|\Phi\rangle \;,(i=1,\ldots,K) \;, \label{keq3}
\end{equation}
and we use the $i$-th expansion above to solve the Hamiltonian $\bar{H}(i)$,
\begin{equation}
\bar{H}(i) = 
\prod_{m=1}^{K/(i)}
e^{-T^{(m)}_{\rm ext}} 
H
\prod_{n=1}^{K/(i)}
e^{T^{(n)}_{\rm ext}}  \;,
\label{mult1}
\end{equation}
where $\prod_{m=1}^{K/(i)} x_m = x_1\ldots x_{i-1}x_{i+1}\ldots x_K$.

Next we follow the same procedure as in the $K=2$ case, i.e., premultiplying the corresponding Schr\"odinger equation by $e^{-T^{(i)}_{\rm ext}}$ and projecting onto $(P+Q_{\rm int})$ we obtain
\begin{equation}
    H^{\rm eff} C_{\rm int}^{(i)}|\Phi\rangle = E_i C_{\rm int}^{(i)}|\Phi\rangle \;, (i=1,\ldots,K) \;,
    \label{hefff}
    \end{equation}
where $\lbrace E_i \rbrace_{i=1}^K$ are  corresponding energies and 
\begin{equation}
H^{\rm eff} = (P+Q_{\rm int}) e^{-\Sigma_{\rm ext}(T^{(1)}_{\rm ext},\ldots ,T^{(K)}_{\rm ext})}
H 
e^{\Sigma_{\rm ext}(T^{(1)}_{\rm ext},\ldots ,T^{(K)}_{\rm ext})}
(P+Q_{\rm int}) \;, \label{uu1}
\end{equation}
where the cumulative external cluster operator for general case is defined as:
\begin{equation}
\Sigma_{\rm ext}(T^{(1)}_{\rm ext},\ldots ,T^{(K)}_{\rm ext}) = \sum_{n=1}^{K} T^{(n)}_{\rm ext} \;. \label{uu2}
\end{equation}
Again, in the proof for general $K$ we used the fact that all external operators commute, $[T^{(i)}_{\rm ext},T^{(j)}_{\rm ext}]=0\;(i,j=1,\ldots,K)$,  and assumed that expansions (\ref{mult1}) refer to the exact case. 
Compared to the bare Hamiltonian $H$ external Hamiltonian,  
$e^{-\Sigma_{\rm ext}(T^{(1)}_{\rm ext},\ldots ,T^{(K)}_{\rm ext})}
H 
e^{\Sigma_{\rm ext}(T^{(1)}_{\rm ext},\ldots ,T^{(K)}_{\rm ext})}$, incorporates external entanglement information (EEI) corresponding to all states of interest.
\qedsymbol{}

The above theorem provides a framework for formulating state-universal effective/downfolded Hamiltonians within the language of single-reference CC theory, while also offering an algorithm to target a selected subset of electronic states non-orthogonal to the reference $|\Phi\rangle$ and approximated by the active space.
The $\Sigma_{\rm ext}$ operator is also a typical example of a multi-purpose cluster operator.
One can envision approximate procedure for calculating truncated forms of $T^{(i)}_{\rm ext}\;, (i=1,\ldots,K)$ using coupled equations equations:
\begin{widetext}
\begin{equation}
    Q_{\rm ext} e^{-\Sigma_{\rm ext}(T^{(1)}_{\rm ext},\ldots ,T^{(K)}_{\rm ext})}
H 
e^{\Sigma_{\rm ext}(T^{(1)}_{\rm ext},\ldots ,T^{(K)}_{\rm ext})} 
\tilde{C}_{\rm int}^{(i)}|\Phi\rangle = 0 \;, (i=1,\ldots,K) \label{uu2} 
\end{equation}
\end{widetext}
where $\tilde{C}_{\rm int}^{(i)}$ is an approximate form of the exact $C_{\rm int}^{(i)}$ and $Q_{\rm ext}$ is the space generated by all Slater determinants generated by acting with exact or approximate $T^{(i)}_{\rm ext}\; (i=1,\ldots,K)$ operators on the reference function $|\Phi\rangle$.  
If all $T^{(i)}_{\rm ext}$ are approximated by the same excitation rank, their numerical identification requires a single computational block that evaluates vectors  ${\bf f}_{\rm ext}(i)$ that corresponds to the left hand side of Eq.~(\ref{uu2}), then the corrections $\Delta {\bf t}^{(i)}_{\rm ext}(n+1)$ to vectors of amplitudes for $i$-th state after $n$ iterations using Newton--Raphson procedure can be given as 
\begin{equation}
    \Delta {\bf t}^{(i)}_{\rm ext}(n+1) \simeq -{\bf J}^{-1} {\bf f}_{\rm ext}(i,n)\;,
    \label{nr1}
\end{equation} 
where ${\bf J}$ is a pre-conditioner approximating the Jacobian matrix of Eq.~(\ref{uu2}. Alternatively, perturbative techniques can be invoked to find approximate forms of the $T_{\rm ext}^{(i)}$ operators.  

Since the similarity-transformed Hamiltonian in (\ref{uu2}), $\bar{H}_{\rm GST}=e^{-\Sigma_{\rm ext}}He^{\Sigma_{\rm ext}}$, depends on $\Sigma_{\rm ext}$, all evaluations of ${\bf f}_{\rm ext}(i)$ differ only in the form of $\tilde{C}_{\rm int}$ the $\bar{H}_{\rm GST}$ is acting on. For this reason we will refer to $\bar{H}_{\rm GST}$ as {\em generating similarity transformed} (GST) Hamiltonian. This observation suggests that the same code can be applied to evaluate corrections (\ref{nr1}), which  naturally supports parallelism of the calculations. 

An advantage of this approach, compared to the genuine multi-reference CC formulation, is that the excitation manifold used in $T^{(i)}_{\rm ext}$ differs significantly from that employed in multi-reference formulations, and the cluster amplitudes defining $T^{(i)}_{\rm ext}$ are governed by large perturbative denominators. In particular, $T^{(i)}_{\rm ext}$ does not involve MR-CC amplitudes, which often lead to the intruder state problem \cite{schucan1972effective,schucan2}. The EEI-guided algorithm for constructing effective Hamiltonians presented here also provides an alternative to multi-reference approaches based on the wave-operator formalism and the Bloch equation \cite{mukherjee1975correlation,jezmonk,lindgren1987connectivity,jeziorski1989valence,meissner1998fock,evangelista2018perspective}.

\section{State-universal-type coupled-cluster downfolding: Hermitian variant}
In the following, using the EEI principle, we extend Theorem 2 to the Hermitian form of effective Hamiltonians. This form not only captures correlation effects for both ground and excited states but also plays a critical role in the early stages of quantum computing, where the number of logical qubits is not yet commensurate with the size of realistic problems defined by the large basis sets required in chemistry and physics to achieve the desired level of accuracy (see, for example, Ref.~\cite{bauman2025coupled}).
%
%
%
\paragraph{Theorem 3 (State-universal Hermitian CC downfolding based on a single-reference unitary CC formalism).---} There exists a single-reference type external anti-Hermitian cluster operator $\Gamma_{\rm ext}$ that defines Hermitian  state-universal effective Hamiltonian, $H^{\rm eff}$, in arbitrary CAS,
$H^{\rm eff} = (P+Q_{\rm int}) e^{-\Gamma_{\rm ext}}He^{\Gamma_{\rm ext}}(P+Q_{\rm int})$ ($\Gamma_{\rm ext}^{\dagger} = -\Gamma_{\rm ext} $), that furnishes energies of arbitrary $K$ states having non-zero overlap with the reference function $|\Phi\rangle$, $1\leq K \leq M$, where $M$ is the dimensionality of the CAS. 
\paragraph{Proof.---} 
In the proof, we will utilize the fact that under certain conditions (associated with the convergence of infinite commutator series originating in multiple use of Baker–Campbell–Hausdorff formula in the context of the  disentangled unitary coupled-cluster Ansatz introduced by Evangelista, Chan, and Scuseria \cite{evangelista2019exact}) the exact wave function $|\Psi\rangle$ takes the form 
\begin{equation}
    |\Psi\rangle = e^{\sigma_{\rm ext}} e^{\sigma_{\rm int}} |\Phi\rangle \;,
    \label{ducc1}
\end{equation}
where anti-Hermitian cluster operators $\sigma_{\rm ext}$ and $\sigma_{\rm int}$ are defined in terms of external and internal SR-type cluster amplitudes (see Ref.~\cite{downfolding2020t} for details).
Alternatively, in analogy to (\ref{adam1}) $|\Psi\rangle$ can be re-represented as 
\begin{equation}
    |\Psi\rangle = e^{\sigma_{\rm ext}} C_{\rm int}|\Phi\rangle \;.
    \label{ducc2}
\end{equation}
Now, let us assume that we have $K$ expansions
\begin{equation}
|\Psi(i)\rangle = e^{\sigma_{\rm ext}^{(i)}} C_{\rm int}^{(i)}|\Phi\rangle \;,(i=1,\ldots,K) \;, \label{ducc3}
\end{equation}
Following the EEI ideas of Theorem 2, we use these expansions to diagonalize Hermitian Hamiltonians
\begin{equation}
\bar{H}(i) = 
\prod_{m=1}^{K/(i)}
e^{-\sigma^{(m)}_{\rm ext}} 
H
\prod_{n=1}^{K/(i)}
e^{\sigma^{(n)}_{\rm ext}}  \;,
\label{hmult1}
\end{equation}
which can be rewritten as 
\begin{equation}
\lbrace 
e^{-\sigma^{(i)}_{\rm ext}}
\prod_{m=1}^{K/(i)}
e^{-\sigma^{(m)}_{\rm ext}}  \rbrace
H
\lbrace 
\lbrack
\prod_{n=1}^{K/(i)}
e^{\sigma^{(n)}_{\rm ext}}
\rbrack
e^{\sigma^{(i)}_{\rm ext}}
\rbrace
e^{\sigma^{(i)}_{\rm int}} |\Phi\rangle = 
E_i 
e^{\sigma^{(i)}_{\rm int}} |\Phi\rangle
\;,
\label{hmult1}
\end{equation}
Using the rank-1 Trotter expansion ($R_T=1$) one can identify the products of exponents above as a single exponent, for example 
$\prod_{n=1}^{K/(i)}
e^{\sigma^{(n)}_{\rm ext}} 
e^{\sigma^{(i)}_{\rm ext}}\simeq
e^{\sum_{m=1}^K \sigma^{(m)}_{\rm ext} }$, which leads to the desired form the Hermitian effective Hamiltonian
\begin{equation}
    H^{\rm eff{}} 
\stackrel{\text{$R_T=1$}}{\simeq}
(P+Q_{\rm int})
e^{-\sum_{m=1}^K \sigma^{(m)}_{\rm ext} } H
e^{\sum_{m=1}^K \sigma^{(m)}_{\rm ext} }
(P+Q_{\rm int}) 
\;,
\label{trotter1}
\end{equation}
which approximately reproduces energies $E_i\;(i=1,\ldots,K)$ and defines cumulative external cluster operator $\Gamma_{\rm ext}$ as $\Gamma_{\rm ext}=\sum_{m=1}^K \sigma^{(m)}_{\rm ext}$. 
However, the rank-1 Trotter formula provides only a basic approximation level. Instead, we would like to have a more rigorous approach that can provide a more accurate/precise form of the effective Hamiltonian. Let us define the unitary operator $U(N,i)$
\begin{equation}
U(N,i) = 
\lbrace
\lbrack
\prod_{n=1}^{K/(i)}
e^{\sigma^{(n)}_{\rm ext}/N} 
\rbrack
e^{\sigma^{(i)}_{\rm ext}/N}
\rbrace^{N-1}
\times
\lbrace 
\prod_{n=1}^{K/(i)}
e^{\sigma^{(n)}_{\rm ext}/N} 
\rbrace 
\;,
\label{trotter2}
\end{equation}
where for the convenience all cluster operators $\sigma^{(k)}_{\rm ext}$ in (\ref{trotter2}) are rescaled by factor $1/N$. Let us also assume that we have also defined an iterative process that expresses external cluster operators for $p+1$-th iteration $\lbrace \sigma_{\rm ext}^{(i)}(p+1)\rbrace_{i=1}^K $ as some functions of these operators in the $p$-th iteration $\lbrace \sigma_{\rm ext}^{(i)}(p)\rbrace_{i=1}^K $. The unitary  operator $U(N,i)$ expressed in terms of $\lbrace \sigma_{\rm ext}^{(i)}(p)\rbrace_{i=1}^K $, we will designate as $U(N,i,p)$. For evaluating values for  $(p+1)$-st iteration of $\sigma_{\rm ext}^{(i)}$  let us assume that the state 
$e^{\sigma_{\rm ext}^{(i)}(p+1)/N}C_{\rm int}^{(i)}(p+1)|\Phi\rangle$ diagonalizes Hamiltonian $\bar{H}(i,p)=U(N,i,p)^{-1}HU(N,i,p)$
that depends on the $\lbrace \sigma_{\rm ext}^{(i)}(p)\rbrace_{i=1}^K $ operators, and the amplitudes defining 
$\sigma_{\rm ext}^{(i)}$ and $C_{\rm int}^{(i)}$ can be 
determined from the variational principle defined by the
functional 
\begin{widetext}
\begin{equation}
\Lambda^{(i)}(p+1)=\langle\Phi|C_{\rm 
int}^{(i)\dagger}(p+1)
e^{-\sigma_{\rm ext}^{(i)}(p+1)/N}\bar{H}
(i,p)e^{\sigma_{\rm ext}^{(i)}(p+1)/N}C_{\rm 
int}^{(i)}(p+1)|\Phi\rangle/\langle\Phi|C_{\rm 
int}^{(i)\dagger}(p+1)
C_{\rm int}^{(i)}(p+1)|\Phi\rangle \;.
\label{u36}
\end{equation}
\end{widetext}
If this process converges to operators $\lbrace \sigma_{\rm ext}^{(i)}\rbrace_{i=1}^K$  and $\lbrace C_{\rm ext}^{(i)}\rbrace_{i=1}^K$ then the solution for $i$-th problem  has to satisfy
\begin{equation}
 U(N,i)^{-1}HU(N,i)e^{\sigma_{\rm ext}^{(i)}/N}C_{\rm 
int}^{(i)}|\Phi\rangle = E_i e^{\sigma_{\rm ext}^{(i)}/N}C_{\rm 
int}^{(i)}|\Phi\rangle \;.
\label{trotter3}
\end{equation}
After  premultiplying (\ref{trotter3}) by $e^{-\sigma_{\rm ext}^{(i)}/N}$, projecting on $(P+Q_{\rm int})$, and noticing that 
\begin{equation}
U(N,i)e^{\sigma_{\rm ext}^{(i)}/N} 
\stackrel{\text{$R_T=N$}}{\simeq}
e^{\sum_{m=1}^K \sigma^{(m)}_{\rm ext} } \;,
\label{trotter4}
\end{equation}
Eq.~(\ref{trotter3}) can be rewritten as 
\begin{equation}
H^{\rm eff} C_{\rm 
int}^{(i)}|\Phi\rangle = E_i C_{\rm 
int}^{(i)}|\Phi\rangle \;,
\label{trotter5}
\end{equation}
where $H^{\rm eff}
\stackrel{\text{$R_T=N$}}{\simeq}
e^{-\Gamma_{\rm ext}}He^{-\Gamma_{\rm ext}}$ and 
$\Gamma_{\rm ext}=\sum_{m=1}^K \sigma^{(m)}_{\rm ext}$
is $i$ independent. Taking $N=\infty$ limit concludes our proof. 
\qedsymbol{}
Theorem 3 paves the way for a multi-state downfolding procedure, in which effective Hamiltonians can be constructed for an arbitrary number of states with non-zero overlap with $|\Phi\rangle$ for arbitrary active spaces. These spaces are chosen, in analogy to standard multi-reference formulations, to strongly overlap with the states of interest. The Hermitian form of state-universal effective Hamiltonians also offers a new approach for simulating excited states on quantum computers. Compared to earlier downfolding techniques for excited states, this method avoids difficulties associated with the state-selective definition of the external operator $\sigma_{\rm ext}$
, which is based on analyzing equation-of-motion CC wave functions, as discussed in Ref. \cite{bauman2019quantum}. This is particularly advantageous for states with symmetries different from that of $|\Phi\rangle$.

In contrast to the non-Hermitian case, Theorem 3 also enables algorithms for  direct  determination of $\Gamma_{\rm ext}$ by optimizing the expression
\begin{equation} 
\min\limits_{\Gamma_{\rm ext},\lbrace C_{\rm int}^{(i)}\rbrace_{i=1}^{K}}
 R = \sum_{i=1}^K \omega_i E^{(i)}(\Gamma_{\rm ext}) \;,
\label{last1}
\end{equation}
where $\lbrace \omega_i \rbrace_{i=1}^K$ are chosen weights and $E^{(i)}$s are eigenvalues of $\Gamma_{\rm ext}$-dependent 
effective Hamiltonian $H^{\rm eff} = (P+Q_{\rm int}) e^{-\Gamma_{\rm ext}}He^{\Gamma_{\rm ext}}(P+Q_{\rm int})$ of Theorem 3.

\section{Conclusion}
In this work, we have identified and analyzed several representative use cases of coupled-cluster (CC) theory in which the single-reference (SR) cluster operator associated with the time-independent Schrödinger equation is partitioned into components that serve distinct and complementary roles. Within this framework, Theorem 1 establishes a general and rigorous mechanism for symmetry breaking in CC theory. When considered in conjunction with the SES theorem, Theorem 1 leads to a rather counterintuitive yet formally exact construction of effective Hamiltonians that recover energies of states belonging to symmetry sectors different from that of the CC reference function. Notably, this construction remains valid even when the active space contains no configurations explicitly associated with the target symmetry.

This result provides a compelling illustration of the universality of the SR-CC ansatz as a vehicle for encoding information about states of different symmetry within a formally single-reference framework. Theorems 2 and 3 further extend this perspective by introducing a many-body algorithm for constructing effective Hamiltonians capable of simultaneously describing multiple states that are non-orthogonal to the reference function. Among these results, Theorem 3 is particularly significant in the context of future quantum simulations of ground and excited states, as it enables a reduced-dimensional representation compatible with limited registers of logical qubits.

Looking forward, the multi-purpose cluster operator framework introduced here naturally admits further generalizations. In particular, ongoing and future extensions will address situations in which the targeted states have zero overlap with the reference function $|\Phi\rangle$, thereby further broadening the scope of SR-CC–based effective Hamiltonian theories for both classical and quantum computing applications.

\section{Acknowledgement}
This material is based upon work supported by the ``Embedding QC into Many-body Frameworks for Strongly Correlated Molecular and Materials Systems''  project, which is funded by the U.S. Department of Energy, Office of Science, Office of Basic Energy Sciences, the Division of Chemical Sciences, Geosciences, and Biosciences (under FWP 72689) and by the Quantum Algorithms and Architecture for Domain Science Initiative (QuAADS), under the Laboratory Directed Research and Development (LDRD) Program at Pacific Northwest National Laboratory (PNNL). This work used resources from the Pacific Northwest National Laboratory (PNNL).
PNNL is operated by Battelle for the U.S. Department of Energy under Contract DE-AC05-76RL01830.





\bibliography{references}

\begin{thebibliography}{60}%
\makeatletter
\providecommand \@ifxundefined [1]{%
 \@ifx{#1\undefined}
}%
\providecommand \@ifnum [1]{%
 \ifnum #1\expandafter \@firstoftwo
 \else \expandafter \@secondoftwo
 \fi
}%
\providecommand \@ifx [1]{%
 \ifx #1\expandafter \@firstoftwo
 \else \expandafter \@secondoftwo
 \fi
}%
\providecommand \natexlab [1]{#1}%
\providecommand \enquote  [1]{``#1''}%
\providecommand \bibnamefont  [1]{#1}%
\providecommand \bibfnamefont [1]{#1}%
\providecommand \citenamefont [1]{#1}%
\providecommand \href@noop [0]{\@secondoftwo}%
\providecommand \href [0]{\begingroup \@sanitize@url \@href}%
\providecommand \@href[1]{\@@startlink{#1}\@@href}%
\providecommand \@@href[1]{\endgroup#1\@@endlink}%
\providecommand \@sanitize@url [0]{\catcode `\\12\catcode `\$12\catcode
  `\&12\catcode `\#12\catcode `\^12\catcode `\_12\catcode `\%12\relax}%
\providecommand \@@startlink[1]{}%
\providecommand \@@endlink[0]{}%
\providecommand \url  [0]{\begingroup\@sanitize@url \@url }%
\providecommand \@url [1]{\endgroup\@href {#1}{\urlprefix }}%
\providecommand \urlprefix  [0]{URL }%
\providecommand \Eprint [0]{\href }%
\providecommand \doibase [0]{http://dx.doi.org/}%
\providecommand \selectlanguage [0]{\@gobble}%
\providecommand \bibinfo  [0]{\@secondoftwo}%
\providecommand \bibfield  [0]{\@secondoftwo}%
\providecommand \translation [1]{[#1]}%
\providecommand \BibitemOpen [0]{}%
\providecommand \bibitemStop [0]{}%
\providecommand \bibitemNoStop [0]{.\EOS\space}%
\providecommand \EOS [0]{\spacefactor3000\relax}%
\providecommand \BibitemShut  [1]{\csname bibitem#1\endcsname}%
\let\auto@bib@innerbib\@empty
\bibitem [{\citenamefont {Coester}(1958)}]{coester58_421}%
  \BibitemOpen
  \bibfield  {author} {\bibinfo {author} {\bibfnamefont {F.}~\bibnamefont
  {Coester}},\ }\bibfield  {title} {\enquote {\bibinfo {title} {Bound states of
  a many-particle system},}\ }\href {\doibase
  http://dx.doi.org/10.1016/0029-5582(58)90280-3} {\bibfield  {journal}
  {\bibinfo  {journal} {Nucl. Phys.}\ }\textbf {\bibinfo {volume} {7}},\
  \bibinfo {pages} {421--424} (\bibinfo {year} {1958})}\BibitemShut {NoStop}%
\bibitem [{\citenamefont {Coester}\ and\ \citenamefont
  {Kummel}(1960)}]{coester60_477}%
  \BibitemOpen
  \bibfield  {author} {\bibinfo {author} {\bibfnamefont {F.}~\bibnamefont
  {Coester}}\ and\ \bibinfo {author} {\bibfnamefont {H.}~\bibnamefont
  {Kummel}},\ }\bibfield  {title} {\enquote {\bibinfo {title} {Short-range
  correlations in nuclear wave functions},}\ }\href {\doibase
  http://dx.doi.org/10.1016/0029-5582(60)90140-1} {\bibfield  {journal}
  {\bibinfo  {journal} {Nucl. Phys.}\ }\textbf {\bibinfo {volume} {17}},\
  \bibinfo {pages} {477--485} (\bibinfo {year} {1960})}\BibitemShut {NoStop}%
\bibitem [{\citenamefont {{\v C}{\'\i}{\v z}ek}(1966)}]{cizek66_4256}%
  \BibitemOpen
  \bibfield  {author} {\bibinfo {author} {\bibfnamefont {J.}~\bibnamefont {{\v
  C}{\'\i}{\v z}ek}},\ }\bibfield  {title} {\enquote {\bibinfo {title} {On the
  correlation problem in atomic and molecular systems. calculation of
  wavefunction components in ursell-type expansion using quantum-field
  theoretical methods},}\ }\href {\doibase http://dx.doi.org/10.1063/1.1727484}
  {\bibfield  {journal} {\bibinfo  {journal} {J. Chem. Phys.}\ }\textbf
  {\bibinfo {volume} {45}},\ \bibinfo {pages} {4256--4266} (\bibinfo {year}
  {1966})}\BibitemShut {NoStop}%
\bibitem [{\citenamefont {Purvis}\ and\ \citenamefont
  {Bartlett}(1982)}]{purvis82_1910}%
  \BibitemOpen
  \bibfield  {author} {\bibinfo {author} {\bibfnamefont {George~D.}\
  \bibnamefont {Purvis}}\ and\ \bibinfo {author} {\bibfnamefont {Rodney~J.}\
  \bibnamefont {Bartlett}},\ }\bibfield  {title} {\enquote {\bibinfo {title} {A
  full coupled-cluster singles and doubles model: The inclusion of disconnected
  triples},}\ }\href {\doibase http://dx.doi.org/10.1063/1.443164} {\bibfield
  {journal} {\bibinfo  {journal} {J. Chem. Phys.}\ }\textbf {\bibinfo {volume}
  {76}},\ \bibinfo {pages} {1910--1918} (\bibinfo {year} {1982})}\BibitemShut
  {NoStop}%
\bibitem [{\citenamefont {Paldus}\ \emph {et~al.}(1972)\citenamefont {Paldus},
  \citenamefont {\ifmmode \check{C}\else
  \v{C}\fi{}\'{\i}\ifmmode~\check{z}\else \v{z}\fi{}ek},\ and\ \citenamefont
  {Shavitt}}]{paldus72_50}%
  \BibitemOpen
  \bibfield  {author} {\bibinfo {author} {\bibfnamefont {J.}~\bibnamefont
  {Paldus}}, \bibinfo {author} {\bibfnamefont {J.}~\bibnamefont {\ifmmode
  \check{C}\else \v{C}\fi{}\'{\i}\ifmmode~\check{z}\else \v{z}\fi{}ek}}, \ and\
  \bibinfo {author} {\bibfnamefont {I.}~\bibnamefont {Shavitt}},\ }\bibfield
  {title} {\enquote {\bibinfo {title} {Correlation problems in atomic and
  molecular systems. iv. extended coupled-pair many-electron theory and its
  application to the b${\mathrm{h}}_{3}$ molecule},}\ }\href {\doibase
  10.1103/PhysRevA.5.50} {\bibfield  {journal} {\bibinfo  {journal} {Phys. Rev.
  A}\ }\textbf {\bibinfo {volume} {5}},\ \bibinfo {pages} {50--67} (\bibinfo
  {year} {1972})}\BibitemShut {NoStop}%
\bibitem [{\citenamefont {Bishop}\ and\ \citenamefont
  {K{\"u}mmel}(1987)}]{bishop1987coupled}%
  \BibitemOpen
  \bibfield  {author} {\bibinfo {author} {\bibfnamefont {Raymond~F}\
  \bibnamefont {Bishop}}\ and\ \bibinfo {author} {\bibfnamefont
  {H}~\bibnamefont {K{\"u}mmel}},\ }\bibfield  {title} {\enquote {\bibinfo
  {title} {The coupled-cluster method},}\ }\href@noop {} {\bibfield  {journal}
  {\bibinfo  {journal} {Phys. Today}\ }\textbf {\bibinfo {volume} {40}},\
  \bibinfo {pages} {52} (\bibinfo {year} {1987})}\BibitemShut {NoStop}%
\bibitem [{\citenamefont {Paldus}\ and\ \citenamefont {Li}(1999)}]{paldus07}%
  \BibitemOpen
  \bibfield  {author} {\bibinfo {author} {\bibfnamefont {Josef}\ \bibnamefont
  {Paldus}}\ and\ \bibinfo {author} {\bibfnamefont {Xiangzhu}\ \bibnamefont
  {Li}},\ }\bibfield  {title} {\enquote {\bibinfo {title} {A critical
  assessment of coupled cluster method in quantum chemistry},}\ }\href
  {\doibase 10.1002/9780470141694.ch1} {\bibfield  {journal} {\bibinfo
  {journal} {Adv. Chem. Phys.}\ }\textbf {\bibinfo {volume} {110}},\ \bibinfo
  {pages} {1--175} (\bibinfo {year} {1999})}\BibitemShut {NoStop}%
\bibitem [{\citenamefont {Bartlett}\ and\ \citenamefont
  {Musia\l{}}(2007)}]{Bartlett2007}%
  \BibitemOpen
  \bibfield  {author} {\bibinfo {author} {\bibfnamefont {R.~J.}\ \bibnamefont
  {Bartlett}}\ and\ \bibinfo {author} {\bibfnamefont {M.}~\bibnamefont
  {Musia\l{}}},\ }\bibfield  {title} {\enquote {\bibinfo {title}
  {Coupled-cluster theory in quantum chemistry},}\ }\href {\doibase
  10.1103/RevModPhys.79.291} {\bibfield  {journal} {\bibinfo  {journal} {Rev.
  Mod. Phys.}\ }\textbf {\bibinfo {volume} {79}},\ \bibinfo {pages} {291--352}
  (\bibinfo {year} {2007})}\BibitemShut {NoStop}%
\bibitem [{\citenamefont {Lyakh}\ \emph {et~al.}(2012)\citenamefont {Lyakh},
  \citenamefont {Musia{\l}}, \citenamefont {Lotrich},\ and\ \citenamefont
  {Bartlett}}]{lyakh2012multireference}%
  \BibitemOpen
  \bibfield  {author} {\bibinfo {author} {\bibfnamefont {Dmitry~I}\
  \bibnamefont {Lyakh}}, \bibinfo {author} {\bibfnamefont {Monika}\
  \bibnamefont {Musia{\l}}}, \bibinfo {author} {\bibfnamefont {Victor~F}\
  \bibnamefont {Lotrich}}, \ and\ \bibinfo {author} {\bibfnamefont {Rodney~J}\
  \bibnamefont {Bartlett}},\ }\bibfield  {title} {\enquote {\bibinfo {title}
  {Multireference nature of chemistry: The coupled-cluster view},}\ }\href@noop
  {} {\bibfield  {journal} {\bibinfo  {journal} {Chemical reviews}\ }\textbf
  {\bibinfo {volume} {112}},\ \bibinfo {pages} {182--243} (\bibinfo {year}
  {2012})}\BibitemShut {NoStop}%
\bibitem [{\citenamefont {Sinha}\ \emph {et~al.}(2012)\citenamefont {Sinha},
  \citenamefont {Maitra},\ and\ \citenamefont
  {Mukherjee}}]{sinha2012development}%
  \BibitemOpen
  \bibfield  {author} {\bibinfo {author} {\bibfnamefont {Debalina}\
  \bibnamefont {Sinha}}, \bibinfo {author} {\bibfnamefont {Rahul}\ \bibnamefont
  {Maitra}}, \ and\ \bibinfo {author} {\bibfnamefont {Debashis}\ \bibnamefont
  {Mukherjee}},\ }\bibfield  {title} {\enquote {\bibinfo {title} {Development
  and applications of a unitary group adapted state specific multi-reference
  coupled cluster theory with internally contracted treatment of inactive
  double excitations},}\ }\href@noop {} {\bibfield  {journal} {\bibinfo
  {journal} {The Journal of Chemical Physics}\ }\textbf {\bibinfo {volume}
  {137}} (\bibinfo {year} {2012})}\BibitemShut {NoStop}%
\bibitem [{\citenamefont {Datta}\ and\ \citenamefont
  {Nooijen}(2012)}]{datta2012multireference}%
  \BibitemOpen
  \bibfield  {author} {\bibinfo {author} {\bibfnamefont {Dipayan}\ \bibnamefont
  {Datta}}\ and\ \bibinfo {author} {\bibfnamefont {Marcel}\ \bibnamefont
  {Nooijen}},\ }\bibfield  {title} {\enquote {\bibinfo {title} {Multireference
  equation-of-motion coupled cluster theory},}\ }\href@noop {} {\bibfield
  {journal} {\bibinfo  {journal} {The Journal of Chemical Physics}\ }\textbf
  {\bibinfo {volume} {137}} (\bibinfo {year} {2012})}\BibitemShut {NoStop}%
\bibitem [{\citenamefont {Musia{\l}}\ \emph {et~al.}(2019)\citenamefont
  {Musia{\l}}, \citenamefont {Meissner},\ and\ \citenamefont
  {Cembrzynska}}]{musial2019intermediate}%
  \BibitemOpen
  \bibfield  {author} {\bibinfo {author} {\bibfnamefont {Monika}\ \bibnamefont
  {Musia{\l}}}, \bibinfo {author} {\bibfnamefont {Leszek}\ \bibnamefont
  {Meissner}}, \ and\ \bibinfo {author} {\bibfnamefont {Justyna}\ \bibnamefont
  {Cembrzynska}},\ }\bibfield  {title} {\enquote {\bibinfo {title} {The
  intermediate hamiltonian fock-space coupled-cluster method with approximate
  evaluation of the three-body effects},}\ }\href@noop {} {\bibfield  {journal}
  {\bibinfo  {journal} {The Journal of Chemical Physics}\ }\textbf {\bibinfo
  {volume} {151}} (\bibinfo {year} {2019})}\BibitemShut {NoStop}%
\bibitem [{\citenamefont {Aoto}\ and\ \citenamefont
  {K{\"o}hn}(2016)}]{aoto2016internally}%
  \BibitemOpen
  \bibfield  {author} {\bibinfo {author} {\bibfnamefont {Yuri~Alexandre}\
  \bibnamefont {Aoto}}\ and\ \bibinfo {author} {\bibfnamefont {Andreas}\
  \bibnamefont {K{\"o}hn}},\ }\bibfield  {title} {\enquote {\bibinfo {title}
  {Internally contracted multireference coupled-cluster theory in a multistate
  framework},}\ }\href@noop {} {\bibfield  {journal} {\bibinfo  {journal} {The
  Journal of Chemical Physics}\ }\textbf {\bibinfo {volume} {144}} (\bibinfo
  {year} {2016})}\BibitemShut {NoStop}%
\bibitem [{\citenamefont {Feldmann}\ and\ \citenamefont
  {Reiher}(2024)}]{feldmann2024renormalized}%
  \BibitemOpen
  \bibfield  {author} {\bibinfo {author} {\bibfnamefont {Robin}\ \bibnamefont
  {Feldmann}}\ and\ \bibinfo {author} {\bibfnamefont {Markus}\ \bibnamefont
  {Reiher}},\ }\bibfield  {title} {\enquote {\bibinfo {title} {Renormalized
  internally contracted multireference coupled cluster with perturbative
  triples},}\ }\href@noop {} {\bibfield  {journal} {\bibinfo  {journal}
  {Journal of Chemical Theory and Computation}\ }\textbf {\bibinfo {volume}
  {20}},\ \bibinfo {pages} {7126--7143} (\bibinfo {year} {2024})}\BibitemShut
  {NoStop}%
\bibitem [{\citenamefont {Adam}\ \emph {et~al.}(2025)\citenamefont {Adam},
  \citenamefont {Waigum},\ and\ \citenamefont
  {K{\"o}hn}}]{adam2025multireference}%
  \BibitemOpen
  \bibfield  {author} {\bibinfo {author} {\bibfnamefont {Robert~G}\
  \bibnamefont {Adam}}, \bibinfo {author} {\bibfnamefont {Alexander}\
  \bibnamefont {Waigum}}, \ and\ \bibinfo {author} {\bibfnamefont {Andreas}\
  \bibnamefont {K{\"o}hn}},\ }\bibfield  {title} {\enquote {\bibinfo {title}
  {Multireference coupled-cluster theory: The internally contracted route},}\
  }\href@noop {} {\bibfield  {journal} {\bibinfo  {journal} {Wiley
  Interdisciplinary Reviews: Computational Molecular Science}\ }\textbf
  {\bibinfo {volume} {15}},\ \bibinfo {pages} {e70023} (\bibinfo {year}
  {2025})}\BibitemShut {NoStop}%
\bibitem [{\citenamefont {Li}\ and\ \citenamefont
  {Evangelista}(2015)}]{li2015multireference}%
  \BibitemOpen
  \bibfield  {author} {\bibinfo {author} {\bibfnamefont {Chenyang}\
  \bibnamefont {Li}}\ and\ \bibinfo {author} {\bibfnamefont {Francesco~A}\
  \bibnamefont {Evangelista}},\ }\bibfield  {title} {\enquote {\bibinfo {title}
  {Multireference driven similarity renormalization group: A second-order
  perturbative analysis},}\ }\href@noop {} {\bibfield  {journal} {\bibinfo
  {journal} {Journal of Chemical Theory and Computation}\ }\textbf {\bibinfo
  {volume} {11}},\ \bibinfo {pages} {2097--2108} (\bibinfo {year}
  {2015})}\BibitemShut {NoStop}%
\bibitem [{\citenamefont {Sokolov}\ and\ \citenamefont
  {Chan}(2016)}]{sokolov2016time}%
  \BibitemOpen
  \bibfield  {author} {\bibinfo {author} {\bibfnamefont {Alexander~Yu}\
  \bibnamefont {Sokolov}}\ and\ \bibinfo {author} {\bibfnamefont {Garnet~Kin}\
  \bibnamefont {Chan}},\ }\bibfield  {title} {\enquote {\bibinfo {title} {A
  time-dependent formulation of multi-reference perturbation theory},}\
  }\href@noop {} {\bibfield  {journal} {\bibinfo  {journal} {The Journal of
  Chemical Physics}\ }\textbf {\bibinfo {volume} {144}} (\bibinfo {year}
  {2016})}\BibitemShut {NoStop}%
\bibitem [{\citenamefont {Lechner}\ \emph {et~al.}(2021)\citenamefont
  {Lechner}, \citenamefont {Izs{\'a}k}, \citenamefont {Nooijen},\ and\
  \citenamefont {Neese}}]{lechner2021perturbative}%
  \BibitemOpen
  \bibfield  {author} {\bibinfo {author} {\bibfnamefont {Marvin~H}\
  \bibnamefont {Lechner}}, \bibinfo {author} {\bibfnamefont {R{\'o}bert}\
  \bibnamefont {Izs{\'a}k}}, \bibinfo {author} {\bibfnamefont {Marcel}\
  \bibnamefont {Nooijen}}, \ and\ \bibinfo {author} {\bibfnamefont {Frank}\
  \bibnamefont {Neese}},\ }\bibfield  {title} {\enquote {\bibinfo {title} {A
  perturbative approach to multireference equation-of-motion coupled
  cluster},}\ }\href@noop {} {\bibfield  {journal} {\bibinfo  {journal}
  {Molecular Physics}\ }\textbf {\bibinfo {volume} {119}},\ \bibinfo {pages}
  {e1939185} (\bibinfo {year} {2021})}\BibitemShut {NoStop}%
\bibitem [{\citenamefont {Li}\ \emph {et~al.}(2023)\citenamefont {Li},
  \citenamefont {Misiewicz},\ and\ \citenamefont
  {Evangelista}}]{li2023intruder}%
  \BibitemOpen
  \bibfield  {author} {\bibinfo {author} {\bibfnamefont {Shuhang}\ \bibnamefont
  {Li}}, \bibinfo {author} {\bibfnamefont {Jonathon~P}\ \bibnamefont
  {Misiewicz}}, \ and\ \bibinfo {author} {\bibfnamefont {Francesco~A}\
  \bibnamefont {Evangelista}},\ }\bibfield  {title} {\enquote {\bibinfo {title}
  {Intruder-free cumulant-truncated driven similarity renormalization group
  second-order multireference perturbation theory},}\ }\href@noop {} {\bibfield
   {journal} {\bibinfo  {journal} {The Journal of Chemical Physics}\ }\textbf
  {\bibinfo {volume} {159}} (\bibinfo {year} {2023})}\BibitemShut {NoStop}%
\bibitem [{\citenamefont {Hayashi}\ \emph {et~al.}(2024)\citenamefont
  {Hayashi}, \citenamefont {Saitow}, \citenamefont {Uemura},\ and\
  \citenamefont {Yanai}}]{hayashi2024quasi}%
  \BibitemOpen
  \bibfield  {author} {\bibinfo {author} {\bibfnamefont {Manami}\ \bibnamefont
  {Hayashi}}, \bibinfo {author} {\bibfnamefont {Masaaki}\ \bibnamefont
  {Saitow}}, \bibinfo {author} {\bibfnamefont {Kazuma}\ \bibnamefont {Uemura}},
  \ and\ \bibinfo {author} {\bibfnamefont {Takeshi}\ \bibnamefont {Yanai}},\
  }\bibfield  {title} {\enquote {\bibinfo {title} {Quasi-degenerate extension
  of local n-electron valence state perturbation theory with pair-natural
  orbital method based on localized virtual molecular orbitals},}\ }\href@noop
  {} {\bibfield  {journal} {\bibinfo  {journal} {The Journal of Chemical
  Physics}\ }\textbf {\bibinfo {volume} {160}} (\bibinfo {year}
  {2024})}\BibitemShut {NoStop}%
\bibitem [{\citenamefont {Abraham}\ \emph {et~al.}(2024)\citenamefont
  {Abraham}, \citenamefont {Atalar}, \citenamefont {Berard}, \citenamefont
  {Booth}, \citenamefont {Burton}, \citenamefont {Chan}, \citenamefont
  {Evangelista}, \citenamefont {Filip}, \citenamefont {Giner}, \citenamefont
  {Gunasekera} \emph {et~al.}}]{abraham2024novel}%
  \BibitemOpen
  \bibfield  {author} {\bibinfo {author} {\bibfnamefont {Vibin}\ \bibnamefont
  {Abraham}}, \bibinfo {author} {\bibfnamefont {Kemal}\ \bibnamefont {Atalar}},
  \bibinfo {author} {\bibfnamefont {Kenneth~O}\ \bibnamefont {Berard}},
  \bibinfo {author} {\bibfnamefont {George~H}\ \bibnamefont {Booth}}, \bibinfo
  {author} {\bibfnamefont {Hugh~GA}\ \bibnamefont {Burton}}, \bibinfo {author}
  {\bibfnamefont {Garnet K-L}\ \bibnamefont {Chan}}, \bibinfo {author}
  {\bibfnamefont {Francesco~A}\ \bibnamefont {Evangelista}}, \bibinfo {author}
  {\bibfnamefont {Maria-Andreea}\ \bibnamefont {Filip}}, \bibinfo {author}
  {\bibfnamefont {Emmanuel}\ \bibnamefont {Giner}}, \bibinfo {author}
  {\bibfnamefont {Alexander}\ \bibnamefont {Gunasekera}},  \emph {et~al.},\
  }\bibfield  {title} {\enquote {\bibinfo {title} {Novel perturbative and
  variational methods for stronger correlations: general discussion},}\
  }\href@noop {} {\bibfield  {journal} {\bibinfo  {journal} {Faraday
  Discussions}\ }\textbf {\bibinfo {volume} {254}},\ \bibinfo {pages}
  {191--215} (\bibinfo {year} {2024})}\BibitemShut {NoStop}%
\bibitem [{\citenamefont {Verma}\ \emph {et~al.}(2025)\citenamefont {Verma},
  \citenamefont {Mitra}, \citenamefont {Wang}, \citenamefont {D’Cunha},
  \citenamefont {Jangid}, \citenamefont {Hennefarth}, \citenamefont {Agarawal},
  \citenamefont {Otis}, \citenamefont {Haldar}, \citenamefont {Hermes} \emph
  {et~al.}}]{verma2025multireference}%
  \BibitemOpen
  \bibfield  {author} {\bibinfo {author} {\bibfnamefont {Shreya}\ \bibnamefont
  {Verma}}, \bibinfo {author} {\bibfnamefont {Abhishek}\ \bibnamefont {Mitra}},
  \bibinfo {author} {\bibfnamefont {Qiaohong}\ \bibnamefont {Wang}}, \bibinfo
  {author} {\bibfnamefont {Ruhee}\ \bibnamefont {D’Cunha}}, \bibinfo {author}
  {\bibfnamefont {Bhavnesh}\ \bibnamefont {Jangid}}, \bibinfo {author}
  {\bibfnamefont {Matthew~R}\ \bibnamefont {Hennefarth}}, \bibinfo {author}
  {\bibfnamefont {Valay}\ \bibnamefont {Agarawal}}, \bibinfo {author}
  {\bibfnamefont {Leon}\ \bibnamefont {Otis}}, \bibinfo {author} {\bibfnamefont
  {Soumi}\ \bibnamefont {Haldar}}, \bibinfo {author} {\bibfnamefont
  {Matthew~R}\ \bibnamefont {Hermes}},  \emph {et~al.},\ }\bibfield  {title}
  {\enquote {\bibinfo {title} {Multireference embedding and fragmentation
  methods for classical and quantum computers: From model systems to realistic
  applications},}\ }\href@noop {} {\bibfield  {journal} {\bibinfo  {journal}
  {Chemical Reviews}\ } (\bibinfo {year} {2025})}\BibitemShut {NoStop}%
\bibitem [{\citenamefont {King}\ \emph {et~al.}(2025)\citenamefont {King},
  \citenamefont {Jangid}, \citenamefont {Hermes},\ and\ \citenamefont
  {Gagliardi}}]{king2025bridging}%
  \BibitemOpen
  \bibfield  {author} {\bibinfo {author} {\bibfnamefont {Daniel~S}\
  \bibnamefont {King}}, \bibinfo {author} {\bibfnamefont {Bhavnesh}\
  \bibnamefont {Jangid}}, \bibinfo {author} {\bibfnamefont {Matthew~R}\
  \bibnamefont {Hermes}}, \ and\ \bibinfo {author} {\bibfnamefont {Laura}\
  \bibnamefont {Gagliardi}},\ }\bibfield  {title} {\enquote {\bibinfo {title}
  {Bridging the gap between molecules and materials in quantum chemistry with
  localized active spaces},}\ }\href@noop {} {\bibfield  {journal} {\bibinfo
  {journal} {Nature Communications}\ }\textbf {\bibinfo {volume} {16}},\
  \bibinfo {pages} {10832} (\bibinfo {year} {2025})}\BibitemShut {NoStop}%
\bibitem [{\citenamefont {Huang}\ \emph {et~al.}(2023)\citenamefont {Huang},
  \citenamefont {Li},\ and\ \citenamefont {Evangelista}}]{huang2023leveraging}%
  \BibitemOpen
  \bibfield  {author} {\bibinfo {author} {\bibfnamefont {Renke}\ \bibnamefont
  {Huang}}, \bibinfo {author} {\bibfnamefont {Chenyang}\ \bibnamefont {Li}}, \
  and\ \bibinfo {author} {\bibfnamefont {Francesco~A}\ \bibnamefont
  {Evangelista}},\ }\bibfield  {title} {\enquote {\bibinfo {title} {Leveraging
  small-scale quantum computers with unitarily downfolded hamiltonians},}\
  }\href@noop {} {\bibfield  {journal} {\bibinfo  {journal} {PRX Quantum}\
  }\textbf {\bibinfo {volume} {4}},\ \bibinfo {pages} {020313} (\bibinfo {year}
  {2023})}\BibitemShut {NoStop}%
\bibitem [{\citenamefont {Lang}\ \emph {et~al.}(2021)\citenamefont {Lang},
  \citenamefont {Ryabinkin},\ and\ \citenamefont {Izmaylov}}]{lang2020unitary}%
  \BibitemOpen
  \bibfield  {author} {\bibinfo {author} {\bibfnamefont {Robert~A.}\
  \bibnamefont {Lang}}, \bibinfo {author} {\bibfnamefont {Ilya~G.}\
  \bibnamefont {Ryabinkin}}, \ and\ \bibinfo {author} {\bibfnamefont
  {Artur~F.}\ \bibnamefont {Izmaylov}},\ }\bibfield  {title} {\enquote
  {\bibinfo {title} {Unitary transformation of the electronic hamiltonian with
  an exact quadratic truncation of the baker-campbell-hausdorff expansion},}\
  }\href@noop {} {\bibfield  {journal} {\bibinfo  {journal} {J. Chem. Theory
  Comput.}\ }\textbf {\bibinfo {volume} {17}},\ \bibinfo {pages} {66--78}
  (\bibinfo {year} {2021})}\BibitemShut {NoStop}%
\bibitem [{\citenamefont {Dhawan}\ \emph {et~al.}(2021)\citenamefont {Dhawan},
  \citenamefont {Metcalf},\ and\ \citenamefont {Zgid}}]{dhawan2021dynamical}%
  \BibitemOpen
  \bibfield  {author} {\bibinfo {author} {\bibfnamefont {Diksha}\ \bibnamefont
  {Dhawan}}, \bibinfo {author} {\bibfnamefont {Mekena}\ \bibnamefont
  {Metcalf}}, \ and\ \bibinfo {author} {\bibfnamefont {Dominika}\ \bibnamefont
  {Zgid}},\ }\bibfield  {title} {\enquote {\bibinfo {title} {Dynamical
  self-energy mapping (dsem) for creation of sparse hamiltonians suitable for
  quantum computing},}\ }\href@noop {} {\bibfield  {journal} {\bibinfo
  {journal} {Journal of chemical theory and computation}\ }\textbf {\bibinfo
  {volume} {17}},\ \bibinfo {pages} {7622--7631} (\bibinfo {year}
  {2021})}\BibitemShut {NoStop}%
\bibitem [{\citenamefont {Kumar}\ \emph {et~al.}(2022)\citenamefont {Kumar},
  \citenamefont {Asthana}, \citenamefont {Masteran}, \citenamefont {Valeev},
  \citenamefont {Zhang}, \citenamefont {Cincio}, \citenamefont {Tretiak},\ and\
  \citenamefont {Dub}}]{kumar2022quantum}%
  \BibitemOpen
  \bibfield  {author} {\bibinfo {author} {\bibfnamefont {Ashutosh}\
  \bibnamefont {Kumar}}, \bibinfo {author} {\bibfnamefont {Ayush}\ \bibnamefont
  {Asthana}}, \bibinfo {author} {\bibfnamefont {Conner}\ \bibnamefont
  {Masteran}}, \bibinfo {author} {\bibfnamefont {Edward~F}\ \bibnamefont
  {Valeev}}, \bibinfo {author} {\bibfnamefont {Yu}~\bibnamefont {Zhang}},
  \bibinfo {author} {\bibfnamefont {Lukasz}\ \bibnamefont {Cincio}}, \bibinfo
  {author} {\bibfnamefont {Sergei}\ \bibnamefont {Tretiak}}, \ and\ \bibinfo
  {author} {\bibfnamefont {Pavel~A}\ \bibnamefont {Dub}},\ }\bibfield  {title}
  {\enquote {\bibinfo {title} {Quantum simulation of molecular electronic
  states with a transcorrelated hamiltonian: higher accuracy with fewer
  qubits},}\ }\href@noop {} {\bibfield  {journal} {\bibinfo  {journal} {Journal
  of chemical theory and computation}\ }\textbf {\bibinfo {volume} {18}},\
  \bibinfo {pages} {5312--5324} (\bibinfo {year} {2022})}\BibitemShut {NoStop}%
\bibitem [{\citenamefont {Kowalski}(2018)}]{kowalski2018properties}%
  \BibitemOpen
  \bibfield  {author} {\bibinfo {author} {\bibfnamefont {Karol}\ \bibnamefont
  {Kowalski}},\ }\bibfield  {title} {\enquote {\bibinfo {title} {Properties of
  coupled-cluster equations originating in excitation sub-algebras},}\
  }\href@noop {} {\bibfield  {journal} {\bibinfo  {journal} {J. Chem. Phys.}\
  }\textbf {\bibinfo {volume} {148}},\ \bibinfo {pages} {094104} (\bibinfo
  {year} {2018})}\BibitemShut {NoStop}%
\bibitem [{\citenamefont {Bauman}\ \emph
  {et~al.}(2019{\natexlab{a}})\citenamefont {Bauman}, \citenamefont {Bylaska},
  \citenamefont {Krishnamoorthy}, \citenamefont {Low}, \citenamefont {Wiebe},
  \citenamefont {Granade}, \citenamefont {Roetteler}, \citenamefont {Troyer},\
  and\ \citenamefont {Kowalski}}]{bauman2019downfolding}%
  \BibitemOpen
  \bibfield  {author} {\bibinfo {author} {\bibfnamefont {N.~P.}\ \bibnamefont
  {Bauman}}, \bibinfo {author} {\bibfnamefont {E.~J.}\ \bibnamefont {Bylaska}},
  \bibinfo {author} {\bibfnamefont {S.}~\bibnamefont {Krishnamoorthy}},
  \bibinfo {author} {\bibfnamefont {G.~H.}\ \bibnamefont {Low}}, \bibinfo
  {author} {\bibfnamefont {N.}~\bibnamefont {Wiebe}}, \bibinfo {author}
  {\bibfnamefont {C.~E.}\ \bibnamefont {Granade}}, \bibinfo {author}
  {\bibfnamefont {M.}~\bibnamefont {Roetteler}}, \bibinfo {author}
  {\bibfnamefont {M.}~\bibnamefont {Troyer}}, \ and\ \bibinfo {author}
  {\bibfnamefont {K.}~\bibnamefont {Kowalski}},\ }\bibfield  {title} {\enquote
  {\bibinfo {title} {Downfolding of many-body hamiltonians using active-space
  models: Extension of the sub-system embedding sub-algebras approach to
  unitary coupled cluster formalisms},}\ }\href {\doibase 10.1063/1.5094643}
  {\bibfield  {journal} {\bibinfo  {journal} {J. Chem. Phys.}\ }\textbf
  {\bibinfo {volume} {151}},\ \bibinfo {pages} {014107} (\bibinfo {year}
  {2019}{\natexlab{a}})}\BibitemShut {NoStop}%
\bibitem [{\citenamefont {Kowalski}(2021)}]{kowalski2021dimensionality}%
  \BibitemOpen
  \bibfield  {author} {\bibinfo {author} {\bibfnamefont {Karol}\ \bibnamefont
  {Kowalski}},\ }\bibfield  {title} {\enquote {\bibinfo {title} {Dimensionality
  reduction of the many-body problem using coupled-cluster subsystem flow
  equations: Classical and quantum computing perspective},}\ }\href@noop {}
  {\bibfield  {journal} {\bibinfo  {journal} {Phys. Rev. A}\ }\textbf {\bibinfo
  {volume} {104}},\ \bibinfo {pages} {032804} (\bibinfo {year}
  {2021})}\BibitemShut {NoStop}%
\bibitem [{\citenamefont {Evangelista}(2014)}]{evangelista2014driven}%
  \BibitemOpen
  \bibfield  {author} {\bibinfo {author} {\bibfnamefont {Francesco~A}\
  \bibnamefont {Evangelista}},\ }\bibfield  {title} {\enquote {\bibinfo {title}
  {A driven similarity renormalization group approach to quantum many-body
  problems},}\ }\href@noop {} {\bibfield  {journal} {\bibinfo  {journal} {The
  Journal of chemical physics}\ }\textbf {\bibinfo {volume} {141}} (\bibinfo
  {year} {2014})}\BibitemShut {NoStop}%
\bibitem [{\citenamefont {Welborn}\ \emph {et~al.}(2016)\citenamefont
  {Welborn}, \citenamefont {Tsuchimochi},\ and\ \citenamefont
  {Van~Voorhis}}]{welborn2016bootstrap}%
  \BibitemOpen
  \bibfield  {author} {\bibinfo {author} {\bibfnamefont {Matthew}\ \bibnamefont
  {Welborn}}, \bibinfo {author} {\bibfnamefont {Takashi}\ \bibnamefont
  {Tsuchimochi}}, \ and\ \bibinfo {author} {\bibfnamefont {Troy}\ \bibnamefont
  {Van~Voorhis}},\ }\bibfield  {title} {\enquote {\bibinfo {title} {Bootstrap
  embedding: An internally consistent fragment-based method},}\ }\href@noop {}
  {\bibfield  {journal} {\bibinfo  {journal} {The Journal of Chemical Physics}\
  }\textbf {\bibinfo {volume} {145}} (\bibinfo {year} {2016})}\BibitemShut
  {NoStop}%
\bibitem [{\citenamefont {Shee}\ \emph {et~al.}(2024)\citenamefont {Shee},
  \citenamefont {Faulstich}, \citenamefont {Whaley}, \citenamefont {Lin},\ and\
  \citenamefont {Head-Gordon}}]{shee2024static}%
  \BibitemOpen
  \bibfield  {author} {\bibinfo {author} {\bibfnamefont {Avijit}\ \bibnamefont
  {Shee}}, \bibinfo {author} {\bibfnamefont {Fabian~M}\ \bibnamefont
  {Faulstich}}, \bibinfo {author} {\bibfnamefont {K~Birgitta}\ \bibnamefont
  {Whaley}}, \bibinfo {author} {\bibfnamefont {Lin}\ \bibnamefont {Lin}}, \
  and\ \bibinfo {author} {\bibfnamefont {Martin}\ \bibnamefont {Head-Gordon}},\
  }\bibfield  {title} {\enquote {\bibinfo {title} {A static quantum embedding
  scheme based on coupled cluster theory},}\ }\href@noop {} {\bibfield
  {journal} {\bibinfo  {journal} {The Journal of Chemical Physics}\ }\textbf
  {\bibinfo {volume} {161}} (\bibinfo {year} {2024})}\BibitemShut {NoStop}%
\bibitem [{\citenamefont {Feldmann}\ \emph {et~al.}(2024)\citenamefont
  {Feldmann}, \citenamefont {M{\"o}rchen}, \citenamefont {Lang}, \citenamefont
  {Lesiuk},\ and\ \citenamefont {Reiher}}]{feldmann2024complete}%
  \BibitemOpen
  \bibfield  {author} {\bibinfo {author} {\bibfnamefont {Robin}\ \bibnamefont
  {Feldmann}}, \bibinfo {author} {\bibfnamefont {Maximilian}\ \bibnamefont
  {M{\"o}rchen}}, \bibinfo {author} {\bibfnamefont {Jakub}\ \bibnamefont
  {Lang}}, \bibinfo {author} {\bibfnamefont {Micha{\l}}\ \bibnamefont
  {Lesiuk}}, \ and\ \bibinfo {author} {\bibfnamefont {Markus}\ \bibnamefont
  {Reiher}},\ }\bibfield  {title} {\enquote {\bibinfo {title} {Complete active
  space iterative coupled cluster theory},}\ }\href@noop {} {\bibfield
  {journal} {\bibinfo  {journal} {The Journal of Physical Chemistry A}\
  }\textbf {\bibinfo {volume} {128}},\ \bibinfo {pages} {8615--8627} (\bibinfo
  {year} {2024})}\BibitemShut {NoStop}%
\bibitem [{\citenamefont {Weisburn}\ \emph {et~al.}(2025)\citenamefont
  {Weisburn}, \citenamefont {Cho}, \citenamefont {Bensberg}, \citenamefont
  {Meitei}, \citenamefont {Reiher},\ and\ \citenamefont
  {Van~Voorhis}}]{weisburn2025multiscale}%
  \BibitemOpen
  \bibfield  {author} {\bibinfo {author} {\bibfnamefont {Leah~P}\ \bibnamefont
  {Weisburn}}, \bibinfo {author} {\bibfnamefont {Minsik}\ \bibnamefont {Cho}},
  \bibinfo {author} {\bibfnamefont {Moritz}\ \bibnamefont {Bensberg}}, \bibinfo
  {author} {\bibfnamefont {Oinam~Romesh}\ \bibnamefont {Meitei}}, \bibinfo
  {author} {\bibfnamefont {Markus}\ \bibnamefont {Reiher}}, \ and\ \bibinfo
  {author} {\bibfnamefont {Troy}\ \bibnamefont {Van~Voorhis}},\ }\bibfield
  {title} {\enquote {\bibinfo {title} {Multiscale embedding for quantum
  computing},}\ }\href@noop {} {\bibfield  {journal} {\bibinfo  {journal}
  {Journal of Chemical Theory and Computation}\ }\textbf {\bibinfo {volume}
  {21}},\ \bibinfo {pages} {4591--4603} (\bibinfo {year} {2025})}\BibitemShut
  {NoStop}%
\bibitem [{\citenamefont {Piecuch}\ \emph {et~al.}(1993)\citenamefont
  {Piecuch}, \citenamefont {Oliphant},\ and\ \citenamefont
  {Adamowicz}}]{pnl93}%
  \BibitemOpen
  \bibfield  {author} {\bibinfo {author} {\bibfnamefont {Piotr}\ \bibnamefont
  {Piecuch}}, \bibinfo {author} {\bibfnamefont {Nevin}\ \bibnamefont
  {Oliphant}}, \ and\ \bibinfo {author} {\bibfnamefont {Ludwik}\ \bibnamefont
  {Adamowicz}},\ }\bibfield  {title} {\enquote {\bibinfo {title} {A
  state-selective multireference coupled-cluster theory employing the
  single-reference formalism},}\ }\href {\doibase 10.1063/1.466179} {\bibfield
  {journal} {\bibinfo  {journal} {J. Chem. Phys.}\ }\textbf {\bibinfo {volume}
  {99}},\ \bibinfo {pages} {1875--1900} (\bibinfo {year} {1993})}\BibitemShut
  {NoStop}%
\bibitem [{\citenamefont {Piecuch}\ and\ \citenamefont
  {Adamowicz}(1994)}]{piecuch1994state}%
  \BibitemOpen
  \bibfield  {author} {\bibinfo {author} {\bibfnamefont {Piotr}\ \bibnamefont
  {Piecuch}}\ and\ \bibinfo {author} {\bibfnamefont {Ludwik}\ \bibnamefont
  {Adamowicz}},\ }\bibfield  {title} {\enquote {\bibinfo {title}
  {State-selective multireference coupled-cluster theory employing the
  single-reference formalism: Implementation and application to the h8 model
  system},}\ }\href@noop {} {\bibfield  {journal} {\bibinfo  {journal} {The
  Journal of chemical physics}\ }\textbf {\bibinfo {volume} {100}},\ \bibinfo
  {pages} {5792--5809} (\bibinfo {year} {1994})}\BibitemShut {NoStop}%
\bibitem [{\citenamefont {Jankowski}\ and\ \citenamefont
  {Kowalski}(1999)}]{jankowski1999physical}%
  \BibitemOpen
  \bibfield  {author} {\bibinfo {author} {\bibfnamefont {K}~\bibnamefont
  {Jankowski}}\ and\ \bibinfo {author} {\bibfnamefont {K}~\bibnamefont
  {Kowalski}},\ }\bibfield  {title} {\enquote {\bibinfo {title} {Physical and
  mathematical content of coupled-cluster equations. ii. on the origin of
  irregular solutions and their elimination via symmetry adaptation},}\
  }\href@noop {} {\bibfield  {journal} {\bibinfo  {journal} {The Journal of
  chemical physics}\ }\textbf {\bibinfo {volume} {110}},\ \bibinfo {pages}
  {9345--9352} (\bibinfo {year} {1999})}\BibitemShut {NoStop}%
\bibitem [{\citenamefont {Mayhall}\ and\ \citenamefont
  {Raghavachari}(2010)}]{mayhall2010multiple}%
  \BibitemOpen
  \bibfield  {author} {\bibinfo {author} {\bibfnamefont {Nicholas~J}\
  \bibnamefont {Mayhall}}\ and\ \bibinfo {author} {\bibfnamefont {Krishnan}\
  \bibnamefont {Raghavachari}},\ }\bibfield  {title} {\enquote {\bibinfo
  {title} {Multiple solutions to the single-reference ccsd equations for
  nih},}\ }\href@noop {} {\bibfield  {journal} {\bibinfo  {journal} {Journal of
  Chemical Theory and Computation}\ }\textbf {\bibinfo {volume} {6}},\ \bibinfo
  {pages} {2714--2720} (\bibinfo {year} {2010})}\BibitemShut {NoStop}%
\bibitem [{\citenamefont {Lee}\ \emph {et~al.}(2019)\citenamefont {Lee},
  \citenamefont {Small},\ and\ \citenamefont {Head-Gordon}}]{lee2019excited}%
  \BibitemOpen
  \bibfield  {author} {\bibinfo {author} {\bibfnamefont {Joonho}\ \bibnamefont
  {Lee}}, \bibinfo {author} {\bibfnamefont {David~W}\ \bibnamefont {Small}}, \
  and\ \bibinfo {author} {\bibfnamefont {Martin}\ \bibnamefont {Head-Gordon}},\
  }\bibfield  {title} {\enquote {\bibinfo {title} {Excited states via coupled
  cluster theory without equation-of-motion methods: Seeking higher roots with
  application to doubly excited states and double core hole states},}\
  }\href@noop {} {\bibfield  {journal} {\bibinfo  {journal} {The Journal of
  chemical physics}\ }\textbf {\bibinfo {volume} {151}} (\bibinfo {year}
  {2019})}\BibitemShut {NoStop}%
\bibitem [{\citenamefont {Faulstich}\ and\ \citenamefont
  {Laestadius}(2023)}]{faulstich2023homotopy}%
  \BibitemOpen
  \bibfield  {author} {\bibinfo {author} {\bibfnamefont {Fabian~M}\
  \bibnamefont {Faulstich}}\ and\ \bibinfo {author} {\bibfnamefont {Andre}\
  \bibnamefont {Laestadius}},\ }\bibfield  {title} {\enquote {\bibinfo {title}
  {Homotopy continuation methods for coupled-cluster theory in quantum
  chemistry},}\ }\href@noop {} {\bibfield  {journal} {\bibinfo  {journal}
  {Molecular Physics}\ ,\ \bibinfo {pages} {e2258599}} (\bibinfo {year}
  {2023})}\BibitemShut {NoStop}%
\bibitem [{\citenamefont {Faulstich}\ and\ \citenamefont
  {Oster}(2024)}]{faulstich2024coupled}%
  \BibitemOpen
  \bibfield  {author} {\bibinfo {author} {\bibfnamefont {Fabian~M}\
  \bibnamefont {Faulstich}}\ and\ \bibinfo {author} {\bibfnamefont {Mathias}\
  \bibnamefont {Oster}},\ }\bibfield  {title} {\enquote {\bibinfo {title}
  {Coupled cluster theory: Toward an algebraic geometry formulation},}\
  }\href@noop {} {\bibfield  {journal} {\bibinfo  {journal} {SIAM Journal on
  Applied Algebra and Geometry}\ }\textbf {\bibinfo {volume} {8}},\ \bibinfo
  {pages} {138--188} (\bibinfo {year} {2024})}\BibitemShut {NoStop}%
\bibitem [{\citenamefont {Sverrisd{\'o}ttir}\ and\ \citenamefont
  {Faulstich}(2024)}]{sverrisdottir2024exploring}%
  \BibitemOpen
  \bibfield  {author} {\bibinfo {author} {\bibfnamefont {Svala}\ \bibnamefont
  {Sverrisd{\'o}ttir}}\ and\ \bibinfo {author} {\bibfnamefont {Fabian~M}\
  \bibnamefont {Faulstich}},\ }\bibfield  {title} {\enquote {\bibinfo {title}
  {Exploring ground and excited states via single reference coupled-cluster
  theory and algebraic geometry},}\ }\href@noop {} {\bibfield  {journal}
  {\bibinfo  {journal} {Journal of Chemical Theory and Computation}\ }\textbf
  {\bibinfo {volume} {20}},\ \bibinfo {pages} {8517--8528} (\bibinfo {year}
  {2024})}\BibitemShut {NoStop}%
\bibitem [{\citenamefont {Liang}\ \emph {et~al.}(2024)\citenamefont {Liang},
  \citenamefont {Kowalski}, \citenamefont {Yang},\ and\ \citenamefont
  {Bauman}}]{liang2024effective}%
  \BibitemOpen
  \bibfield  {author} {\bibinfo {author} {\bibfnamefont {Senwei}\ \bibnamefont
  {Liang}}, \bibinfo {author} {\bibfnamefont {Karol}\ \bibnamefont {Kowalski}},
  \bibinfo {author} {\bibfnamefont {Chao}\ \bibnamefont {Yang}}, \ and\
  \bibinfo {author} {\bibfnamefont {Nicholas~P}\ \bibnamefont {Bauman}},\
  }\bibfield  {title} {\enquote {\bibinfo {title} {Effective many-body
  interactions in reduced-dimensionality spaces through neural network
  models},}\ }\href@noop {} {\bibfield  {journal} {\bibinfo  {journal}
  {Physical Review Research}\ }\textbf {\bibinfo {volume} {6}},\ \bibinfo
  {pages} {043287} (\bibinfo {year} {2024})}\BibitemShut {NoStop}%
\bibitem [{\citenamefont {Bauman}\ \emph {et~al.}(2025)\citenamefont {Bauman},
  \citenamefont {Zheng}, \citenamefont {Liu}, \citenamefont {Myers},
  \citenamefont {Panyala}, \citenamefont {Peng}, \citenamefont {Li},\ and\
  \citenamefont {Kowalski}}]{bauman2025coupled}%
  \BibitemOpen
  \bibfield  {author} {\bibinfo {author} {\bibfnamefont {Nicholas~P}\
  \bibnamefont {Bauman}}, \bibinfo {author} {\bibfnamefont {Muqing}\
  \bibnamefont {Zheng}}, \bibinfo {author} {\bibfnamefont {Chenxu}\
  \bibnamefont {Liu}}, \bibinfo {author} {\bibfnamefont {Nathan~M}\
  \bibnamefont {Myers}}, \bibinfo {author} {\bibfnamefont {Ajay}\ \bibnamefont
  {Panyala}}, \bibinfo {author} {\bibfnamefont {Bo}~\bibnamefont {Peng}},
  \bibinfo {author} {\bibfnamefont {Ang}\ \bibnamefont {Li}}, \ and\ \bibinfo
  {author} {\bibfnamefont {Karol}\ \bibnamefont {Kowalski}},\ }\bibfield
  {title} {\enquote {\bibinfo {title} {Coupled cluster downfolding theoryin
  simulations of chemical systems on quantum hardware},}\ }\href@noop {}
  {\bibfield  {journal} {\bibinfo  {journal} {arXiv preprint arXiv:2507.01199}\
  } (\bibinfo {year} {2025})}\BibitemShut {NoStop}%
\bibitem [{\citenamefont {Kowalski}\ and\ \citenamefont
  {Bauman}(2020)}]{downfolding2020t}%
  \BibitemOpen
  \bibfield  {author} {\bibinfo {author} {\bibfnamefont {Karol}\ \bibnamefont
  {Kowalski}}\ and\ \bibinfo {author} {\bibfnamefont {Nicholas~P.}\
  \bibnamefont {Bauman}},\ }\bibfield  {title} {\enquote {\bibinfo {title}
  {Sub-system quantum dynamics using coupled cluster downfolding techniques},}\
  }\href {\doibase 10.1063/5.0008436} {\bibfield  {journal} {\bibinfo
  {journal} {J. Chem. Phys.}\ }\textbf {\bibinfo {volume} {152}},\ \bibinfo
  {pages} {244127} (\bibinfo {year} {2020})}\BibitemShut {NoStop}%
\bibitem [{\citenamefont {Ivanov}\ and\ \citenamefont
  {Adamowicz}(2000{\natexlab{a}})}]{ivanov2000casccd}%
  \BibitemOpen
  \bibfield  {author} {\bibinfo {author} {\bibfnamefont {Vladimir~V}\
  \bibnamefont {Ivanov}}\ and\ \bibinfo {author} {\bibfnamefont {Ludwik}\
  \bibnamefont {Adamowicz}},\ }\bibfield  {title} {\enquote {\bibinfo {title}
  {Casccd: Coupled-cluster method with double excitations and the cas
  reference},}\ }\href@noop {} {\bibfield  {journal} {\bibinfo  {journal} {The
  Journal of Chemical Physics}\ }\textbf {\bibinfo {volume} {112}},\ \bibinfo
  {pages} {9258--9268} (\bibinfo {year} {2000}{\natexlab{a}})}\BibitemShut
  {NoStop}%
\bibitem [{\citenamefont {Ivanov}\ and\ \citenamefont
  {Adamowicz}(2000{\natexlab{b}})}]{ivanov2000new}%
  \BibitemOpen
  \bibfield  {author} {\bibinfo {author} {\bibfnamefont {Vladimir~V}\
  \bibnamefont {Ivanov}}\ and\ \bibinfo {author} {\bibfnamefont {Ludwik}\
  \bibnamefont {Adamowicz}},\ }\bibfield  {title} {\enquote {\bibinfo {title}
  {New scheme for solving the amplitude equations in the state-specific coupled
  cluster theory with complete active space reference for ground and excited
  states},}\ }\href@noop {} {\bibfield  {journal} {\bibinfo  {journal} {The
  Journal of Chemical Physics}\ }\textbf {\bibinfo {volume} {113}},\ \bibinfo
  {pages} {8503--8513} (\bibinfo {year} {2000}{\natexlab{b}})}\BibitemShut
  {NoStop}%
\bibitem [{\citenamefont {Adamowicz}\ \emph {et~al.}(2000)\citenamefont
  {Adamowicz}, \citenamefont {Malrieu},\ and\ \citenamefont
  {Ivanov}}]{adamowicz2000new}%
  \BibitemOpen
  \bibfield  {author} {\bibinfo {author} {\bibfnamefont {Ludwik}\ \bibnamefont
  {Adamowicz}}, \bibinfo {author} {\bibfnamefont {Jean-Paul}\ \bibnamefont
  {Malrieu}}, \ and\ \bibinfo {author} {\bibfnamefont {Vladimir~V}\
  \bibnamefont {Ivanov}},\ }\bibfield  {title} {\enquote {\bibinfo {title} {New
  approach to the state-specific multireference coupled-cluster formalism},}\
  }\href@noop {} {\bibfield  {journal} {\bibinfo  {journal} {The Journal of
  Chemical Physics}\ }\textbf {\bibinfo {volume} {112}},\ \bibinfo {pages}
  {10075--10084} (\bibinfo {year} {2000})}\BibitemShut {NoStop}%
\bibitem [{\citenamefont {Ivanov}\ \emph {et~al.}(2009)\citenamefont {Ivanov},
  \citenamefont {Lyakh},\ and\ \citenamefont
  {Adamowicz}}]{ivanov2009multireference}%
  \BibitemOpen
  \bibfield  {author} {\bibinfo {author} {\bibfnamefont {Vladimir~V}\
  \bibnamefont {Ivanov}}, \bibinfo {author} {\bibfnamefont {Dmitry~I}\
  \bibnamefont {Lyakh}}, \ and\ \bibinfo {author} {\bibfnamefont {Ludwik}\
  \bibnamefont {Adamowicz}},\ }\bibfield  {title} {\enquote {\bibinfo {title}
  {Multireference state-specific coupled-cluster methods. state-of-the-art and
  perspectives},}\ }\href@noop {} {\bibfield  {journal} {\bibinfo  {journal}
  {Physical Chemistry Chemical Physics}\ }\textbf {\bibinfo {volume} {11}},\
  \bibinfo {pages} {2355--2370} (\bibinfo {year} {2009})}\BibitemShut {NoStop}%
\bibitem [{\citenamefont {Schucan}\ and\ \citenamefont
  {Weidenm{\"u}ller}(1972)}]{schucan1972effective}%
  \BibitemOpen
  \bibfield  {author} {\bibinfo {author} {\bibfnamefont {TH}~\bibnamefont
  {Schucan}}\ and\ \bibinfo {author} {\bibfnamefont {HA}~\bibnamefont
  {Weidenm{\"u}ller}},\ }\bibfield  {title} {\enquote {\bibinfo {title} {The
  effective interaction in nuclei and its perturbation expansion: An algebraic
  approach},}\ }\href@noop {} {\bibfield  {journal} {\bibinfo  {journal} {Ann.
  Phys.}\ }\textbf {\bibinfo {volume} {73}},\ \bibinfo {pages} {108--135}
  (\bibinfo {year} {1972})}\BibitemShut {NoStop}%
\bibitem [{\citenamefont {Schucan}\ and\ \citenamefont
  {Weidenm{\"u}eller}(1973)}]{schucan2}%
  \BibitemOpen
  \bibfield  {author} {\bibinfo {author} {\bibfnamefont {Thomas~H}\
  \bibnamefont {Schucan}}\ and\ \bibinfo {author} {\bibfnamefont {Hans~A}\
  \bibnamefont {Weidenm{\"u}eller}},\ }\bibfield  {title} {\enquote {\bibinfo
  {title} {Perturbation theory for the effective interaction in nuclei},}\
  }\href {\doibase https://doi.org/10.1016/0003-4916(73)90044-4} {\bibfield
  {journal} {\bibinfo  {journal} {Ann. Phys.}\ }\textbf {\bibinfo {volume}
  {76}},\ \bibinfo {pages} {483--509} (\bibinfo {year} {1973})}\BibitemShut
  {NoStop}%
\bibitem [{\citenamefont {Mukherjee}\ \emph {et~al.}(1975)\citenamefont
  {Mukherjee}, \citenamefont {Moitra},\ and\ \citenamefont
  {Mukhopadhyay}}]{mukherjee1975correlation}%
  \BibitemOpen
  \bibfield  {author} {\bibinfo {author} {\bibfnamefont {Debashis}\
  \bibnamefont {Mukherjee}}, \bibinfo {author} {\bibfnamefont {Raj~Kumar}\
  \bibnamefont {Moitra}}, \ and\ \bibinfo {author} {\bibfnamefont {Atri}\
  \bibnamefont {Mukhopadhyay}},\ }\bibfield  {title} {\enquote {\bibinfo
  {title} {Correlation problem in open-shell atoms and molecules: A
  non-perturbative linked cluster formulation},}\ }\href@noop {} {\bibfield
  {journal} {\bibinfo  {journal} {Molecular Physics}\ }\textbf {\bibinfo
  {volume} {30}},\ \bibinfo {pages} {1861--1888} (\bibinfo {year}
  {1975})}\BibitemShut {NoStop}%
\bibitem [{\citenamefont {Jeziorski}\ and\ \citenamefont
  {Monkhorst}(1981)}]{jezmonk}%
  \BibitemOpen
  \bibfield  {author} {\bibinfo {author} {\bibfnamefont {Bogumil}\ \bibnamefont
  {Jeziorski}}\ and\ \bibinfo {author} {\bibfnamefont {Hendrik~J.}\
  \bibnamefont {Monkhorst}},\ }\bibfield  {title} {\enquote {\bibinfo {title}
  {Coupled-cluster method for multideterminantal reference states},}\ }\href
  {\doibase 10.1103/PhysRevA.24.1668} {\bibfield  {journal} {\bibinfo
  {journal} {Phys. Rev. A}\ }\textbf {\bibinfo {volume} {24}},\ \bibinfo
  {pages} {1668--1681} (\bibinfo {year} {1981})}\BibitemShut {NoStop}%
\bibitem [{\citenamefont {Lindgren}\ and\ \citenamefont
  {Mukherjee}(1987)}]{lindgren1987connectivity}%
  \BibitemOpen
  \bibfield  {author} {\bibinfo {author} {\bibfnamefont {Ingvar}\ \bibnamefont
  {Lindgren}}\ and\ \bibinfo {author} {\bibfnamefont {Debashis}\ \bibnamefont
  {Mukherjee}},\ }\bibfield  {title} {\enquote {\bibinfo {title} {On the
  connectivity criteria in the open-shell coupled-cluster theory for general
  model spaces},}\ }\href@noop {} {\bibfield  {journal} {\bibinfo  {journal}
  {Physics Reports}\ }\textbf {\bibinfo {volume} {151}},\ \bibinfo {pages}
  {93--127} (\bibinfo {year} {1987})}\BibitemShut {NoStop}%
\bibitem [{\citenamefont {Jeziorski}\ and\ \citenamefont
  {Paldus}(1989)}]{jeziorski1989valence}%
  \BibitemOpen
  \bibfield  {author} {\bibinfo {author} {\bibfnamefont {Bogumil}\ \bibnamefont
  {Jeziorski}}\ and\ \bibinfo {author} {\bibfnamefont {Josef}\ \bibnamefont
  {Paldus}},\ }\bibfield  {title} {\enquote {\bibinfo {title} {Valence
  universal exponential ansatz and the cluster structure of multireference
  configuration interaction wave function},}\ }\href@noop {} {\bibfield
  {journal} {\bibinfo  {journal} {The Journal of chemical physics}\ }\textbf
  {\bibinfo {volume} {90}},\ \bibinfo {pages} {2714--2731} (\bibinfo {year}
  {1989})}\BibitemShut {NoStop}%
\bibitem [{\citenamefont {Meissner}(1998)}]{meissner1998fock}%
  \BibitemOpen
  \bibfield  {author} {\bibinfo {author} {\bibfnamefont {Leszek}\ \bibnamefont
  {Meissner}},\ }\bibfield  {title} {\enquote {\bibinfo {title} {Fock-space
  coupled-cluster method in the intermediate hamiltonian formulation: Model
  with singles and doubles},}\ }\href@noop {} {\bibfield  {journal} {\bibinfo
  {journal} {The Journal of chemical physics}\ }\textbf {\bibinfo {volume}
  {108}},\ \bibinfo {pages} {9227--9235} (\bibinfo {year} {1998})}\BibitemShut
  {NoStop}%
\bibitem [{\citenamefont {Evangelista}(2018)}]{evangelista2018perspective}%
  \BibitemOpen
  \bibfield  {author} {\bibinfo {author} {\bibfnamefont {Francesco~A}\
  \bibnamefont {Evangelista}},\ }\bibfield  {title} {\enquote {\bibinfo {title}
  {Perspective: Multireference coupled cluster theories of dynamical electron
  correlation},}\ }\href@noop {} {\bibfield  {journal} {\bibinfo  {journal}
  {The Journal of Chemical Physics}\ }\textbf {\bibinfo {volume} {149}}
  (\bibinfo {year} {2018})}\BibitemShut {NoStop}%
\bibitem [{\citenamefont {Evangelista}\ \emph {et~al.}(2019)\citenamefont
  {Evangelista}, \citenamefont {Chan},\ and\ \citenamefont
  {Scuseria}}]{evangelista2019exact}%
  \BibitemOpen
  \bibfield  {author} {\bibinfo {author} {\bibfnamefont {Francesco~A}\
  \bibnamefont {Evangelista}}, \bibinfo {author} {\bibfnamefont {Garnet
  Kin-Lic}\ \bibnamefont {Chan}}, \ and\ \bibinfo {author} {\bibfnamefont
  {Gustavo~E}\ \bibnamefont {Scuseria}},\ }\bibfield  {title} {\enquote
  {\bibinfo {title} {Exact parameterization of fermionic wave functions via
  unitary coupled cluster theory},}\ }\href@noop {} {\bibfield  {journal}
  {\bibinfo  {journal} {J. Chem. Phys.}\ }\textbf {\bibinfo {volume} {151}},\
  \bibinfo {pages} {244112} (\bibinfo {year} {2019})}\BibitemShut {NoStop}%
\bibitem [{\citenamefont {Bauman}\ \emph
  {et~al.}(2019{\natexlab{b}})\citenamefont {Bauman}, \citenamefont {Low},\
  and\ \citenamefont {Kowalski}}]{bauman2019quantum}%
  \BibitemOpen
  \bibfield  {author} {\bibinfo {author} {\bibfnamefont {Nicholas~P}\
  \bibnamefont {Bauman}}, \bibinfo {author} {\bibfnamefont {Guang~Hao}\
  \bibnamefont {Low}}, \ and\ \bibinfo {author} {\bibfnamefont {Karol}\
  \bibnamefont {Kowalski}},\ }\bibfield  {title} {\enquote {\bibinfo {title}
  {Quantum simulations of excited states with active-space downfolded
  hamiltonians},}\ }\href@noop {} {\bibfield  {journal} {\bibinfo  {journal}
  {The Journal of chemical physics}\ }\textbf {\bibinfo {volume} {151}}
  (\bibinfo {year} {2019}{\natexlab{b}})}\BibitemShut {NoStop}%
\end{thebibliography}%

\end{document}